\documentclass[onecolumn,12pt,twoside]{IEEEtran}
\ifCLASSINFOpdf
 \usepackage[pdftex]{graphicx}
\else
  \usepackage[dvips]{graphicx}
\fi

\usepackage[cmex10]{amsmath}
\usepackage{amsfonts}
\usepackage{amssymb}
\usepackage{amsthm}
\usepackage{relsize}
\usepackage{float}
\usepackage{morefloats}
\usepackage{cite}
\usepackage{bm}
\usepackage{url}
\usepackage{setspace}
\usepackage[caption=false,font=normalsize,labelfont=sf,textfont=sf]{subfig}
\usepackage{booktabs}
\theoremstyle{plain}

\begin{document}
%

\bibliographystyle{ieeetran}
\title{Maximum Sum Rate of Slotted Aloha with Successive Interference Cancellation}
\author{Yitong Li and Lin~Dai,~\IEEEmembership{Senior Member,~IEEE}

\thanks{The authors are with the Department
of Electronic Engineering, City University of Hong Kong,
83 Tat Chee Avenue, Kowloon Tong, Hong Kong, China (email: yitongli2-c@my.cityu.edu.hk, lindai@cityu.edu.hk).}}

\maketitle

\begin{abstract}
This is a sequel of our previous work \cite{Dai_capture} on characterization of maximum sum rate of slotted Aloha networks. By extending the analysis to incorporate the capacity-achieving receiver structure, Successive Interference Cancellation (SIC), this paper aims to identify the rate loss due to random access. Specifically, two representative SIC receivers are considered, i.e, ordered SIC where packets are decoded in a descending order of their received power, and unordered SIC where packets are decoded in a random order. The maximum sum rate and the corresponding optimal parameter setting including the transmission probability and the information encoding rate in both cases are obtained as functions of the mean received signal-to-noise ratio (SNR). The comparison to the capture model shows that the gains are significant only with the ordered SIC at moderate values of the mean received SNR $\rho$. With a large $\rho$, the rate gap diminishes, and they all have the same high-SNR slope of $e^{-1}$, which is far below that of the ergodic sum capacity of fading channels. The effect of multipacket reception (MPR) on the sum rate performance is also studied by comparing the MPR receivers including SIC and the capture model to the classical collision model.

\end{abstract}


\section{Introduction}\label{Section1}

As a basic type of multiple access, random access has been widely applied to a plethora of networks including cellular networks, IEEE 802.11 networks and wireless ad-hoc networks \cite{Kurose,Murthy,Ghavimi}. In sharp contrast to its simplicity in concept and success in applications, however, the theory of random access is far from complete \cite{Minero}. One fundamental question that demands an answer is, for instance: What is the maximum information-theoretic sum rate of random access, and how far is it from the sum capacity of multiple access? This paper aims to contribute to the understanding of the above open question by specifically focusing on slotted Aloha \cite{Abramson}, the simplest and one of the most representative random-access schemes.

Let us start by summarizing two basic features that are shared by most random-access schemes.
\begin{enumerate}
\item \textit{Independent Encoding and Uncoordinated Transmissions}:  Each node independently encodes its information, and decides when to transmit. As the subset of active nodes is random and time-varying, it is normally assumed to be unknown at the transmitter side.

\item \textit{Time-Slotted and Packet-Based}: Though it may not seem to be essential or necessary for random access, a time-slotted and packet-based network has been assumed in the literature since Abramson's Aloha was proposed \cite{Abramson}. Due to the uncoordinated transmissions of nodes, not all the transmitted packets can be successfully decoded in each time slot.
\end{enumerate}

With the above features of random access, the number of successfully decoded packets would be varying from time to time. As a result, most studies have focused on the average number of successfully decoded packets per time slot, which is referred to as the network throughput. To analyze the information-theoretic sum rate of a time-slotted packet-based random-access network, a key assumption was further made in \cite{Angel} that each packet consists of one codeword, and the codeword length is long enough such that a packet can be successfully decoded as long as its information encoding rate does not exceed its single-user capacity, i.e., $\log_2 (1+\eta)$, where $\eta$ is the signal-to-interference-plus-noise ratio (SINR) of the packet. Moreover, it is commonly assumed that all the nodes have a uniform information encoding rate if their channel statistics are identical and they are unaware of the instantaneous realization of the channel gains \cite{Minero,Angel,Naware1}. Under these assumptions, the sum rate of a random-access network is determined by both the encoding rate of each packet and the network throughput.

As demonstrated in our recent work \cite{Dai_capture}, there exists an inherent tradeoff between the encoding rate and the network throughput, which has been largely ignored in previous studies. Intuitively, for given channel statistics, the higher encoding rate of each packet, the fewer packets that can be successfully decoded on average, and thus the lower network throughput. As a result, the sum rate can be further maximized by optimizing the information encoding rate of each packet. In \cite{Dai_capture}, we characterized the maximum sum rate of an $n$-node slotted Aloha network over Rayleigh fading channels with the capture model at the receiver side, that is, each packet is decoded by treating others as background noise, and is successfully decoded if its SINR is above a certain threshold. Both the maximum sum rate and the optimal setting including the SINR threshold
and the transmission probabilities of nodes are obtained as functions of the mean received signal-to-noise ratio (SNR) and the number of nodes. The Additive-White-Gaussian-Noise (AWGN) channels were further considered in \cite{Li_CISS}, and the comparison to \cite{Dai_capture} corroborates that the maximum sum rate of Aloha with the capture model in the fading case would be significantly lower than that with AWGN channels if the channel state information is not available at the transmitter side.

In this paper, the analysis will further be extended to incorporate a more sophisticated type of receivers, Successive Interference Cancellation (SIC) \cite{Cover}. The reason why we consider SIC is two-fold. First of all, it has been shown in \cite{Dai_capture, Li_CISS} that the high-SNR slope of the maximum sum rate of slotted Aloha with the capture model is only $e^{-1}$, which is far below that of the ergodic sum capacity of multiple access fading channels. It is, nevertheless, difficult to determine whether the gap is caused by the random access nature or the suboptimality of the receiver, i.e., the capture model. By adopting the capacity-achieving receiver structure \cite{Tse}, we can pinpoint the rate loss due to random access. Moreover, SIC and the capture model both have the so-called multipacket reception (MPR) capability \cite{Ghez1}. By comparing the MPR receivers including SIC and the capture model to the classical collision model \cite{Abramson}, we can further evaluate the effect of MPR on the maximum sum rate performance of slotted Aloha.

Specifically, we focus on an $n$-node slotted Aloha network where all the nodes transmit to a single receiver\footnote{The multiple access scenario considered in this paper should be distinguished from the ad-hoc scenario where multiple transmitter-receiver pairs exist \cite{Bacelli_2006,Weber_2010}.} with the SINR threshold $\mu$, and the received SNRs of nodes' packets are assumed to be exponentially distributed with the mean received SNR $\rho$. Two representative SIC receivers are considered, i.e., ordered SIC where packets are decoded in a descending order of their received power, and unordered SIC where packets are decoded in a random order. The maximum network throughput and the corresponding optimal transmission probability of nodes in both cases are derived as functions of the SINR threshold $\mu$ and the mean received SNR $\rho$. The maximum sum rate is further obtained by optimizing the SINR threshold $\mu$, and the comparison to the capture model \cite{Dai_capture} shows that ordering is crucial for the rate performance of slotted Aloha with SIC receivers: Without proper ordering of packets, the maximum sum rate could be degraded to that with the capture model.

To demonstrate the effect of MPR on the sum rate performance, the maximum sum rates with SIC receivers and the capture model are further compared to that with the classical collision model. It is found that in contrast to the significant throughput improvement at the low SNR region, only marginal gains in the maximum sum rate can be achieved by SIC receivers and the capture model. The rate difference becomes negligible when the mean received SNR $\rho$ is large, and the maximum sum rates with all the receivers have the same high-SNR slope of $e^{-1}$, which is remarkably lower than that of the ergodic sum capacity of multiple access fading channels.

The remainder of this paper is organized as follows. Section \ref{Section2-R} presents a detailed review of the related work on random access with MPR. The system model is introduced in Section \ref{Section2}. The network steady-state point in saturated conditions and the maximum sum rate are characterized in Section \ref{Section3} and Section \ref{Section4}, respectively. The effect of MPR on the sum rate performance of slotted Aloha and the rate loss due to random access are discussed in Section \ref{Section5}. Conclusions are summarized in Section \ref{Section6}. Table I lists the main notations used in this paper.

\begin{table*}[!tp]\small\label{Table:main_notation}
  \begin{center}
    \caption{Main Notations}
    \begin{tabular}{cl}
      \toprule
      $n$ & Number of nodes \\
      $\rho$ & Mean received SNR \\
      $\mu$ & SINR threshold\\
      $R$ & Information encoding rate of nodes\\
      $\hat{\lambda}_{out}$ & Network throughput\\
      $r_i$ &Conditional probability of successful transmission of each packet given that there are $i$ concurrent packet transmissions\\
      $p$ & Steady-state probability of successful transmission of each packet\\
      $K$ & Cutoff phase of each packet \\
      $\{q_i\}_{i=0, \ldots, K}$ & Transmission probabilities of nodes \\
      $\hat{\lambda}_{\max}$ & Maximum network throughput\\
      $C$ & Maximum sum rate\\
      \bottomrule
    \end{tabular}
  \end{center}
  \vspace{-5mm}
\end{table*}

\section{Related Work} \label{Section2-R}

There has been a rich literature on the performance analysis of random-access networks. In this section, we limit our discussion to random access with MPR.

Early studies on Aloha networks have been based on the classical collision model \cite{Abramson,Carleial,Lam,Tsybakov,Rao}, which assumes that a packet transmission is successful if and only if there are no concurrent transmissions. Though a useful simplification, with the collision model, the network throughput performance could be greatly underestimated when MPR is possible. Therefore, a more general model was proposed in \cite{Ghez1}, where the MPR capability was described by a reception matrix whose entry in the $k$-th row and $l$-th column $\epsilon_{kl}$ is the conditional probability that $l$ packets are successfully decoded given that there are totally $k$ packet transmissions. Given the reception matrix, the stability regions and network throughput performance of slotted Aloha networks were further studied in \cite{Ghez1,Naware2}.

As the reception matrix depends on the receiver design, various MPR receiver models have been considered in the literature. For instance, it was assumed in \cite{Zhang2009,Zhang2010,chan2013,Bae2014} that the receiver can successfully decode up to $M$ packets in each time slot, that is, $\epsilon_{kl}=1$ if $l\leq M \leq k$ and $\epsilon_{kl}=0$ otherwise. Here $M$ represents the MPR capability of the receiver, and its effects on the network throughput performance were analyzed in \cite{Zhang2009,Zhang2010,chan2013,Bae2014}. For many receiver structures, however, its MPR capability may not be simplified as a fixed number. With the widely-adopted capture model \cite{Roberts,Arnbak,Zorzi,Luo2,Rasool,Dua}, for instance,
the reception probability $\epsilon_{kl}$ becomes a function of the SINR threshold \cite{Zorzi}. In general, the reception matrix $\{\epsilon_{kl}\}$ is determined by the specific receiver structure and the minimum required SINR for successfully decoding a packet, which could be difficult to obtain in some cases. It was shown in \cite{Zanella} that with a multi-stage SIC\footnote{Specifically, it was assumed in \cite{Zanella} that in each stage, the packets with SINR higher than a given threshold are decoded, and subtracted from the total received signal. The process is repeated sequentially until no packets can be successfully decoded or the number of stages reaches the maximum.}, the calculation of the reception probability $\epsilon_{kl}$ involves multiple nested integrals. To reduce the computational complexity, approximations were further proposed in \cite{Zanella}, based on which a simple recursive algorithm was designed to estimate the network throughput.

Though providing a general description for MPR capability, the reception matrix is indeed unnecessary for throughput analysis of random-access networks. Instead of deriving $\epsilon_{kl}$, which could become intractable when sophisticated receiver structures are adopted, the analysis can be established based on the conditional probability of \textit{one packet} being successfully decoded given that it has $i$ concurrent packet transmissions, $r_i$. As we will show in this paper, for an $n$-node saturated slotted Aloha network, both the network throughput and the sum rate can be obtained as explicit functions of $\{r_i\}_{i=0,\ldots,n-1}$.

Note that SIC receivers were also considered in \cite{chongbin2013,Yim,Liang_2017,Shirvanimoghaddam_2017_2,Choi_2017} for various random-access networks under distinct assumptions on power control. Specifically, it was assumed in \cite{chongbin2013} that each user randomly chooses a transmission power level based on a certain distribution. In \cite{Yim,Liang_2017,Shirvanimoghaddam_2017_2,Choi_2017}, a limited number of power levels were assumed to be adopted at the receiver side, where each user adjusts its transmission power based on the channel state information to ensure that its received power belongs to a predefined set.\footnote{Note that a single received power level was assumed in \cite{Shirvanimoghaddam_2017_2}, that is, all the users have the same received power.} The focus has been mainly on evaluation of throughput performance. Yet little attention is paid to the sum rate characterization, as well as how to maximize the sum rate by optimally choosing the transmission probabilities and information encoding rate of each node. Moreover, it is worth mentioning that for the line of research on ``coded random access'' \cite{Casini_2007,Liva_2011,Stefanovic_2013,Goseling_2015,Paolini_2015}, the SIC structure is used to recover the coded successive packets of each node, and the classical collision model is assumed for packets from different nodes. It should be distinguished from this paper where no coding/decoding is assumed over successive packets of each node, and the SIC receiver is adopted for packets from multiple nodes.

\section{System Model}\label{Section2}
Consider a slotted Aloha network where $n$ nodes transmit to a single receiver. All the nodes are synchronized and can start a transmission only at the beginning of a time slot. For each node, assume that it always has packets in its buffer and each packet transmission lasts for one time slot. We assume perfect and instant acknowledgement from the receiver and ignore the subtleties of the physical layer such as the switching time from receiving mode to transmitting mode.

\subsection{Channel Model}\label{Section2-1}
Let $g_k$ denote the channel gain from node $k$ to the receiver, which can be written as $g_k=\gamma_k\cdot h_k$, where $h_k$ is the small-scale fading coefficient of node $k$ that varies from time slot to time slot and is modeled as a complex Gaussian random variable with zero mean and unit variance. The large-scale fading coefficient $\gamma_k$ characterizes the long-term channel effect such as path loss and shadowing. Due to the slow-varying nature, the large-scale fading coefficients are usually available at the transmitter side through channel measurement and feedback.

It was shown in \cite{Dai_capture} that for Aloha networks with the capture model, nodes with higher \textit{mean} received power would have better throughput performance, resulting in unfairness among nodes. Therefore, in this paper, we follow the assumption that the transmission power of each node is properly adjusted to overcome the effect of large-scale fading. That is,
\begin{equation}\label{power_control}
\bar{P}_k \cdot |\gamma_k|^2=P_0,
\end{equation}
where $\bar{P}_k$ denotes the transmission power of node $k$. In this case, each node has the same mean received signal-to-noise ratio (SNR) $\rho=P_0/\sigma^2$. Note that uplink power control has been widely adopted in practical networks such as cellular systems, where the base station sends a pilot signal periodically for all the users in its cell to measure their large-scale fading gains and adjust their transmission power accordingly to maintain constant mean received power.\footnote{Note that similar to \cite{Dai_capture}, power control was assumed in this paper to ensure fairness among nodes, that is, all the nodes have the same \textit{mean} received SNR, and thus the same rate performance. It was recently shown in \cite{Sun_CSMA} that for a multi-rate CSMA network where nodes can have distinct transmission rates, the tradeoff between the maximum sum rate and fairness could be significant when the difference in the transmission rates of nodes is large. For Aloha networks, it can be expected that the maximum sum rate can also be improved by allowing certain unfairness among nodes. Yet how to characterize the rate-fairness tradeoff for multi-rate Aloha networks with SIC receivers is a challenging issue that deserves much attention in the future study.}


\subsection{Packet-based Encoding/Decoding}\label{Section2-2}

As nodes in random-access networks normally do not coordinate their transmissions, we assume that no rate allocation among nodes is performed. Moreover, assume that nodes are unaware of the instantaneous realizations of the small-scale fading coefficients. As a result, each node independently encodes its information at a constant rate $R$ bit/s/Hz. Assume that each packet consists of one codeword, and each codeword lasts for one time slot, i.e., no coding over successive packets.

For each packet, denote its received signal-to-interference-plus-noise ratio (SINR) as $\eta$. For given encoding rate $R$, if $\log_2(1+{\eta})>R  $, then by random coding the error probability of the packet is exponentially reduced to zero as the codeword length goes to infinity. Similar to \cite{Angel} and \cite{Dai_capture}, we assume that the codeword length of each packet is sufficiently large such that a packet can be successfully decoded as long as ${\eta}\geq 2^R-1 $. Denote
\begin{equation}\label{mu}
\mu=2^R-1
\end{equation}
as the SINR threshold. For each packet, if its received SINR exceeds the threshold $\mu$, then it can be successfully decoded and rate $R$ can be supported for reliable communications.

\begin{figure*}[!tp]
\centering
\includegraphics[width=5.8in,height=2.1in]{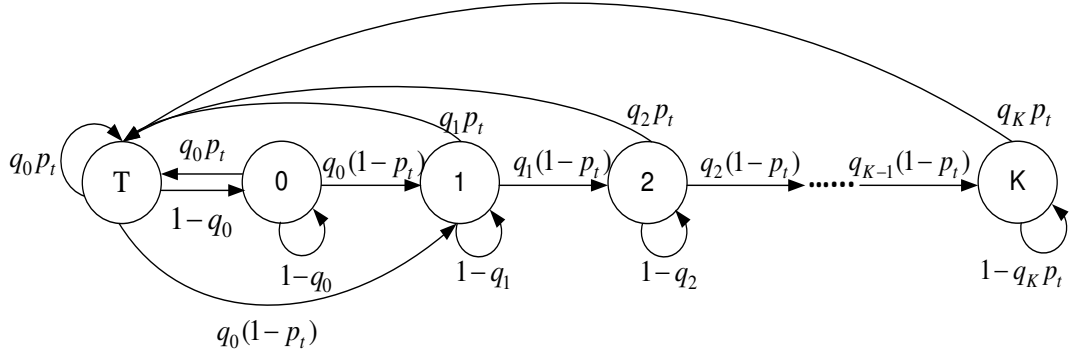}
\caption{State transition diagram of an individual HOL packet in slotted Aloha networks.}\setlength{\abovecaptionskip}{-10cm}
\label{markov}
\end{figure*}

\subsection{Transmitter/Receiver Model}\label{Section2-3}
An $n$-node buffered slotted Aloha network is essentially an $n$-queue-single-server system whose performance is determined by the aggregate activities of head-of-line (HOL) packets. The behavior of each HOL packet can be modeled as a discrete-time Markov process shown in Fig. \ref{markov}. Specifically, a fresh HOL packet is initially in State T, and moves to State 0 if it is not transmitted. Define the phase of a HOL packet as the number of unsuccessful transmissions it experiences. A phase-$i$ HOL packet has a transmission probability of $q_i$, $i=0,\ldots,K$. It stays in State $i$ if it is not transmitted. Otherwise, it moves to State T if its transmission is successful, or State $\min(K,i+1)$ if the transmission fails, where $K$ denotes the cutoff phase. Note that the cutoff phase $K$ can be any non-negative integer. When $K=0$, States 0 and $K$ in Fig. \ref{markov} would be merged into one state, i.e., State 0.

In Fig. \ref{markov}, $p_t$ denotes the probability of successful transmission of HOL packets at time slot $t$. The steady-state probability distribution of the Markov chain in Fig. \ref{markov} can be further obtained as
\begin{equation}
\pi_T=\tfrac{1}{\sum_{i=0}^{K-1}\tfrac{\left(1-p\right)^i}{q_i}+\tfrac{\left(1-p\right)^K}{pq_K}},
\label{pi0}
\end{equation}
and
\begin{equation}
\begin{cases}
\pi_0=\tfrac{1-pq_0}{pq_0}\pi_T.  & K=0 \\
\pi_0=\tfrac{1-q_0}{q_0}\pi_T, \;\;\; \pi_i=\tfrac{(1-p)^i}{q_i}\pi_T, \;i=1,\ldots,K-1, \;\;\; \pi_K=\tfrac{(1-p)^K}{pq_K}\pi_T.  & K\geq 1,
\end{cases}
\label{piK}
\end{equation}
where $p=\mathop{\lim }\limits_{t\to \infty } p_{t}$. Note that $\pi_T$ is the service rate of each node's queue as the queue has a successful output if and only if the HOL packet is in State T.


At the receiver side,
assume that successive interference cancellation (SIC) is adopted, that is, once a packet is successfully decoded, it is subtracted from the aggregate received signal before decoding other packets. There are many variants of the SIC receiver. In this paper, we consider the following two representative cases.

1) Ordered SIC: The most widely adopted SIC assumes that packets are decoded in a descending order of the received power. In each iteration, the packet with the highest received power is decoded, and is subtracted from the aggregate received signal if it is successfully decoded. Otherwise, the decoding process terminates\footnote{If the packet with the highest received power cannot be successfully decoded, then other packets cannot be successfully decoded either.}.

2) Unordered SIC: We also consider the case that packets are decoded in a random order. In each iteration, a packet is decoded and is subtracted from the aggregate received signal if it is successfully decoded. Otherwise, move to the next packet and repeat the process until all the packets are decoded. Here we include the unordered SIC as a comparison benchmark to the ordered SIC as we aim to identify when the rate gain brought by ordering is significant for Aloha networks.

\subsection{Network Throughput and Sum Rate}\label{Section2-4}
In this paper, we focus on the long-term system behavior and define the sum rate as the time-average of the received information rate: $R_s=\lim_{T\to\infty} \frac{1}{T}\sum_{t=1}^{T} R\cdot N_t$, where $N_t$ denotes the number of successfully decoded packets in time slot $t$, which is a time-varying variable. Define the network throughput as the average number of successfully decoded packets per time slot, which can be written as $\hat{\lambda}_{out}=\lim_{T\to\infty} \frac{1}{T}\sum_{t=1}^{T}N_t$. We then have
\begin{equation}\label{sum_rate}
R_s=R \cdot \hat{\lambda}_{out}.
\end{equation}
Note that the sum rate is closely dependent on the transmission probabilities $\{ q_{i} \} _{i=0,...,K} $ and the SINR threshold $\mu$. In this paper, we aim at maximizing the sum rate by optimally choosing the transmission probabilities $\{ q_{i} \} _{i=0,...,K}$ and the SINR threshold $\mu$: $C=\max_{\mu,\{q_i\}}R_s$. By combining \eqref{mu} and \eqref{sum_rate}, the maximum sum rate can be further written as
\begin{equation}\label{maxRate_1}
C=\max_{\mu}\hat{\lambda}_{\max}\log_2(1+\mu),
\end{equation}
where $\hat{\lambda}_{\max}=\max_{\{q_i\}}\hat{\lambda}_{out}$ denotes the maximum network throughput.

For the sake of simplicity, in this paper, we assume that $q_i=q_0$ for $i=0,\ldots,K$. As we can see from \eqref{pi0}, the network throughput in saturated conditions\footnote{To be more specific, the network is saturated if every node always has packets in its buffer. In this paper, we consider the saturated conditions as we aim to obtain the maximum network throughput. In this case, the network throughput is equal to the aggregate service rate of nodes' queues.}, which is determined by the aggregate service rate, closely depends on the steady-state probability of successful transmission of HOL packets $p$. Therefore, before we derive the maximum sum rate in Section \ref{Section4}, let us first focus on the steady-state probability of successful transmission of HOL packets $p$ in the next section.

\section{Steady-State Point in Saturated Conditions}\label{Section3}
In this section, we will characterize the network steady-state point in saturated conditions based on the fixed-point equation of $p$. We will start from the derivation of the conditional probability of successful transmission of each HOL packet given that it has $i$ concurrent packet transmissions, $r_i$.

\subsection{Conditional Probability of Successful Transmission $r_i$}\label{Section3-1}
For HOL packet $j$, denote $r_i^j$ as its steady-state probability of successful transmission given that there are $i$ concurrent packet transmissions. It is clear that $r_i^j$ depends on the receiver design. In \cite{Dai_capture}, it has been shown that with the capture model, all the packets have the same conditional probability of successful transmission of
\begin{equation}\label{ric}
r_i^C=\frac{\exp\left(-\frac{\mu}{\rho}\right)}{(1+\mu)^i},
\end{equation}
where $\mu$ is the SINR threshold and $\rho$ is the mean received SNR.

With SIC, in contrast, $r_i^j$ depends on the decoding order of packet $j$, and the number of packets that have been successfully decoded before packet $j$. Specifically, it can be written as
\begin{align}\label{rij_expre}
r_i^j=&\sum_{m=1}^{i+1}\text{Pr\{packet $j$ is decoded at the $m$th iteration\}}{\cdot}
\sum_{k=0}^{m-1}\text{Pr\{There are $k$ successfully decoded} \notag\\
&\text{packets before packet $j$ and packet $j$ is successfully decoded\}}.
\end{align}
Note that all the packets have independent and identically distributed (i.i.d.) received SNRs, and thus equal probability to be decoded at a given iteration. The first item on the right-hand side of \eqref{rij_expre} is then given by
\begin{equation}\label{rij_expre_1}
\text{Pr\{packet $j$ is decoded at the $m$th iteration\}}=\dfrac{1}{i+1}.
\end{equation}
The second item on the right-hand side of \eqref{rij_expre} depends on whether the packets are ordered. In the following, let us consider the ordered SIC and unordered SIC cases separately.

\subsubsection{Ordered SIC}\label{Section3-1-1}
The second item on the right-hand side of \eqref{rij_expre} can be written as
\begin{align}\label{rij_expre_2_ordered}
&\text{Pr\{There are $k$ successfully decoded packets before packet $j$ and packet $j$ is successfully decoded\}}\notag\\
=&\text{Pr\{There are $k$ successfully decoded packets before packet $j$ $\mid$ packet $j$ is successfully decoded\}}\cdot \notag\\
&\text{Pr\{Packet $j$ is successfully decoded\}}.
\end{align}
With ordered SIC, packets are decoded in a descending order of their received power. If packet $j$ is successfully decoded at the $m$th iteration, then all the $m-1$ packets before packet $j$ must be successfully decoded. Therefore, we have
\begin{align}\label{rij_expre_2_ordered_1}
&\text{Pr\{There are $k$ successfully decoded packets before packet $j$ $\mid$ packet $j$ is successfully decoded\}}\notag\\
&=\begin{cases}
1 & \text{if} \;\;k=m-1, \\
0 & \text{otherwise}.
\end{cases}
\end{align}
Moreover, denote $y_i^{(l)}$ as the probability that the packet at the $l$th iteration can be successfully decoded given that there are $i$ concurrent packet transmissions, i.e., totally $i+1$ packets to be decoded. The probability that packet $j$ is successfully decoded is equal to the probability that all the packets before it, i.e., at iteration $1$ to $m$, are successfully decoded, which can be written as
\begin{equation}\label{rij_expre_2_ordered_2}
\text{Pr\{Packet $j$ is successfully decoded\}}=\Pi_{l=1}^m y_i^{(l)}.
\end{equation}
\noindent Appendix \ref{Appendix_A} further shows that
\begin{equation}\label{y_i^l} y_i^{(l)}=\begin{cases}
\dfrac{(i{+}1)!}{(i{+}1{-}l)!(l{-}1)!}\displaystyle \sum_{k=0}^{\left\lceil1/\mu\right\rceil{-}1} \binom{i{+}1{-}l}{k}\dfrac{(-1)^k}{l{+}k}\exp\left(-\dfrac{l{+}k}{\rho\left(\frac{1}{\mu}{-}k\right)}\right)\left(\dfrac{\frac{1}{\mu}{-}k}{l{+}\frac{1}{\mu}}\right)^{i{+}1{-}l} \;\;l\leq i{+}1{-}\frac{1}{\mu}\\
\\
\dfrac{(i{+}1)!}{(i{+}1{-}l)!(l{-}1)!} \displaystyle\sum_{k=0}^{i{+}1{-}l} \binom{i{+}1{-}l}{k}(-1)^k \left(  \dfrac{\exp\left({-}\frac{k{+}l}{\rho(\frac{1}{\mu}{-}k)}\right)}{k{+}l}-\displaystyle\sum_{s=0}^{i{-}l}\left( \dfrac{(i{+}1{-}l{-}k)^s}{(i{+}1)^{s{+}1}}\cdot \right.\right. \\
Q\left(1{+}s,\dfrac{i{+}1}{\rho(\frac{1}{\mu}{-}i{-}1{+}l)}\right){+}\exp\left({-}\frac{k{+}l}{\rho(\frac{1}{\mu}{-}k)}\right)\dfrac{(\frac{1}{\mu}{-}k)^s}{(\frac{1}{\mu}{+}l)^{s+1}}{\cdot}\left(\vphantom{Q\left(1{+}s,{-}\dfrac{1}{\rho(i{+}1{-}l{-}\frac{1}{\mu})}{\cdot}\dfrac{(i{+}1{-}l{-}k)(\frac{1}{\mu}{+}l)}{\frac{1}{\mu}{-}k} \right)}1{-}\right.\\
\left.\left.\left.\!Q\left(1{+}s,{-}\dfrac{1}{\rho(i{+}1{-}l{-}\frac{1}{\mu})}{\cdot}\dfrac{(i{+}1{-}l{-}k)(\frac{1}{\mu}{+}l)}{\frac{1}{\mu}{-}k} \right)\right)   \right) \vphantom{\dfrac{\exp\left({-}\frac{k{+}l}{\rho(\frac{1}{\mu}{-}k)}\right)}{k{+}l}} \right)\;\;\;\;\;\;\;\;\;\;\;\;\;\;\;\;\;\;\;\;\;\;\;\;\;\;\;\;\;\;\;\;\;\;l>i{+}1{-}\dfrac{1}{\mu},

\end{cases}\end{equation}
for $l=1,...,i+1$, in which $Q(s,t)=\frac{1}{(s-1)!}\int_t^\infty e^{-x}x^{s-1}dx$ is the regularized upper incomplete gamma function. With $\mu\geq1$, \eqref{y_i^l} reduces to
\begin{equation}\label{Mu>1}
y_i^{(l),\mu\geq1}=\binom{i{+}1}{l}\exp\left(-\dfrac{l\mu}{\rho}\right)\cdot\left(\dfrac{1}{l\mu+1}\right)^{i{+}1-l},
\end{equation}
for $l=1,...,i+1$. By substituting (\ref{rij_expre_1})-(\ref{rij_expre_2_ordered_2}) into \eqref{rij_expre}, the steady-state probability of successful transmission of HOL packet $j$ given that there are $i$ concurrent transmissions for ordered SIC can be obtained as
\begin{equation}\label{rij_so}
r_i^{j,OS}=\dfrac{1}{i+1}\sum_{m=1}^{i+1}\Pi_{l=1}^m y_i^{(l)}.
\end{equation}
The right-hand side of \eqref{rij_so} is independent of $j$, indicating that all the HOL packets have the same conditional probability of successful transmission. Therefore, we drop the superscript $j$ in the following.

\subsubsection{Unordered SIC}\label{Section3-1-2}
The second item on the right-hand side of \eqref{rij_expre} can be written as
\begin{align}\label{rij_expre_2_unordered}
&\text{Pr\{There are $k$ successfully decoded packets before packet $j$ and packet $j$ is successfully decoded\}} \notag\\
=&\text{Pr\{Packet $j$ is successfully decoded $\mid$ there are $k$ successfully decoded packets before packet $j$ \}} \cdot \notag\\
&\text{Pr\{There are $k$ successfully decoded packets before packet $j$\}}.
\end{align}
With unordered SIC, packets are decoded in a random order. Therefore, given that there are $k$ successfully decoded packets before packet $j$, the probability that packet $j$ can be successfully decoded is equal to $r_{i-k}^C$, i.e., the probability of successful transmission given that there are $i-k$ concurrent packet transmissions for the capture model. According to \eqref{ric}, we have
\begin{align}\label{rij_expre_2_unordered_1}
&\text{Pr\{Packet $j$ is successfully decoded $\mid$ there are $k$ successfully decoded packets before packet $j$\}}\notag\\
=&\dfrac{\exp\left(-\frac{\mu}{\rho}\right)}{(1+\mu)^{i-k}}=(1+\mu)^k\cdot r_i^C.
\end{align}
It is difficult to obtain an explicit expression of Pr\{There are $k$ successfully decoded packets before packet $j$\} due to the dependency of packets decoding, i.e., the decoding of each packet depends on how many packets at the preceding iterations are successfully decoded. Let us consider the worst case that each packet is subject to interference from $i$ concurrent packet transmissions during its decoding process. In this case, the probability that there are $k$ successfully decoded packets before packet $j$ at the $m$th iteration is given by $\binom{m{-}1}{k}\left(1-r_i^C\right)^{m-1-k}\left(r_i^C\right)^k$. By combining (\ref{rij_expre})-(\ref{rij_expre_1}) and (\ref{rij_expre_2_unordered})-(\ref{rij_expre_2_unordered_1}), we can then obtain a lower-bound of the steady-state probability of successful transmission of HOL packet $j$ given that there are $i$ concurrent packet transmissions for unordered SIC as
\begin{equation}\label{rij_sn}
r_i^{j,NS}\geq r_i^{j,NS\_l}= \tfrac{1}{i+1}\sum_{m=1}^{i+1}\sum_{k=0}^{m-1}\binom{m{-}1}{k}(1+\mu)^k \left(1-r_i^C\right)^{m-1-k}\left(r_i^C\right)^{k+1}=\tfrac{(1+\mu r_i^C)^{i+1}-1}{(i+1)\mu}.
\end{equation}
Again, we drop the superscript $j$ in the following discussion because the right-hand side of \eqref{rij_sn} is independent of $j$.

\subsubsection{Comparison}\label{Section3-3}
\begin{figure*}[!tp]
\centering
\includegraphics[width=4.2in,height=2.5in]{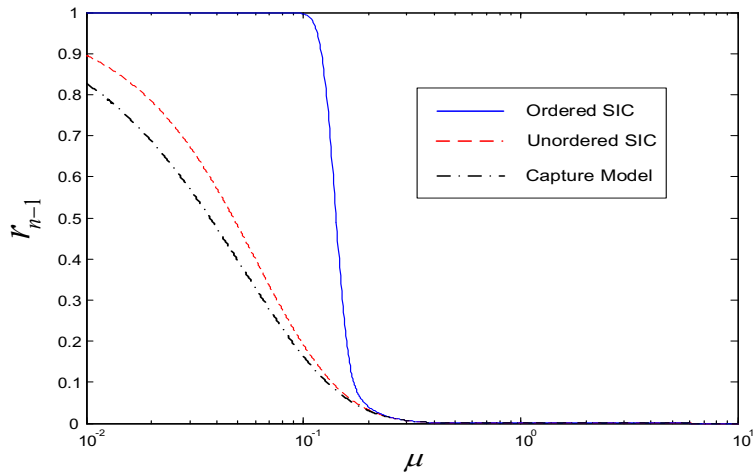}
\caption{Conditional probability of successful transmission $r_{n-1}$ given that there are $n-1$ concurrent packet transmissions versus SINR threshold $\mu$. $n=20$ and $\rho=20$dB.}
\label{ri_mu}
\end{figure*}

We can see from \eqref{rij_so} and \eqref{rij_sn} that both $r_i^{OS}$ and $r_i^{NS\_l}$ depend on the SINR threshold $\mu$. Fig. \ref{ri_mu} illustrates how they vary with $\mu$. For the sake of comparison, the result of the capture model \cite{Dai_capture} is also presented.

It can be clearly observed from Fig. \ref{ri_mu} that both the ordered SIC and unordered SIC have a much larger conditional probability of successful transmission than the capture model when $\mu$ is small. It can be easily obtained from \eqref{ric} that with the capture model, $r_{n-1}^{C}|_{\mu=\frac{1}{n}}\stackrel{\text{for large } n}{\approx}e^{-1}$. Appendix \ref{Proof_Corollary_smallthreshold} further shows that for large $\rho$ and $n$,
\begin{equation}\label{Corollary_smallthreshold}
r_{n-1}^{OS}|_{\mu=\frac{1}{n}}{\approx} 1, \;\;\;\;\; r_{n-1}^{NS\_l}|_{\mu=\frac{1}{n}} {\approx} e^{e^{-1}}-1,
\end{equation}
both of which are significantly higher than $e^{-1}$.

The gap, nevertheless, diminishes as the SINR threshold $\mu$ increases. As $\mu\rightarrow\infty$, Appendix \ref{Proof_Corollary_largethreshold_ri_unordered} shows that the conditional probability of successful transmission in both the ordered SIC and unordered SIC cases converges to that of the capture model.
\begin{equation}\label{Corollary_largethreshold_ri_unordered}
\lim_{\mu\rightarrow\infty}\tfrac{r_i^{OS}}{r_i^C}=\lim_{\mu\rightarrow\infty}\tfrac{r_i^{NS\_l}}{r_i^C}=1.
\end{equation}

\subsection{Steady-State Point in Saturated Conditions}\label{Section3-4}

\begin{figure*}[!tp]
\centering
\includegraphics[width=4.2in,height=2.5in]{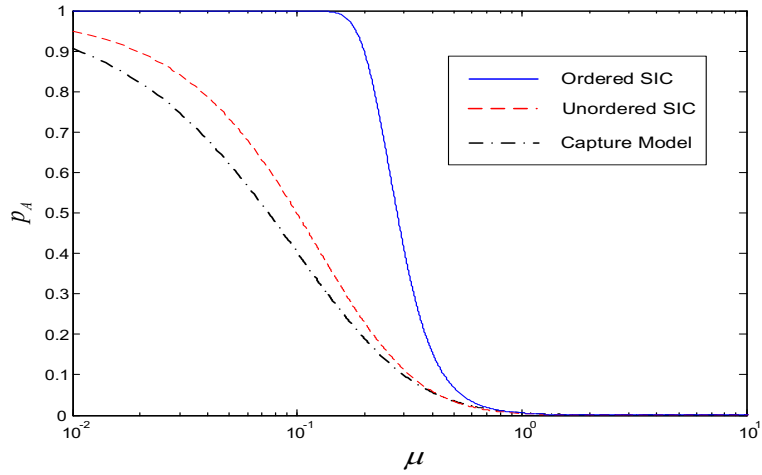}
\caption{Steady-state point $p_A$ versus SINR threshold $\mu$. $n=20$. $q_0=0.5$ and $\rho=20$dB.}
\label{pA_mu}
\end{figure*}

The steady-state probability of successful transmission of HOL packets $p$ can be written as
\begin{equation}\label{Probability-of-success_1}
p=\sum_{i=0}^{n-1} r_i \cdot \text{Pr\{$i$ concurrent packet transmissions\}}.
\end{equation}
In saturated conditions, all the nodes have non-empty queues. According to the Markov chain shown in Fig. \ref{markov}, the probability that the HOL packet is requesting transmission is given by $\pi_T q_0+\sum_{i=0}^K \pi_i q_i$, which is equal to $\pi_T/p$ according to (\ref{pi0})-(\ref{piK}). Therefore, the probability that there are $i$ concurrent packet transmissions can be obtained as
\begin{align}\label{Probability-of-success_2}
\text{Pr\{$i$ concurrent packet transmissions\}}{=}\binom{n{-}1}{i}\left(1{-}\pi_T/p\right)^{n{-}1{-}i}\left(\pi_T/p\right)^{i}.
\end{align}
By substituting \eqref{Probability-of-success_2} into \eqref{Probability-of-success_1}, we have
\begin{equation}\label{Probability-of-success_3}
p=\sum_{i=0}^{n-1} r_i \cdot \binom{n{-}1}{i}\left(1{-}\pi_T/p\right)^{n{-}1{-}i}\left(\pi_T/p\right)^{i}.
\end{equation}
The service rate $\pi_T$ depends on the transmission probabilities of nodes $\{ q_{i} \} _{i=0,...,K}$. With $q_i=q_0$, $i=0,..., K$, the fixed-point equation \eqref{Probability-of-success_3} has a single non-zero root $p_A$, which is given by
\begin{equation}\label{pA_K0}
p_A=\sum_{i=0}^{n-1}\binom{n-1}{i} r_i \left(1-q_0\right)^{n-1-i}\cdot q_0^{i}.
\end{equation}
Specifically, with ordered SIC, $p_A^{OS}$ can be obtained by combining \eqref{pA_K0} and \eqref{rij_so}. With unordered SIC, $p_A^{NS}$ can be approximated by combining \eqref{pA_K0} and the lower-bound $r_i^{NS\_l}$ developed in \eqref{rij_sn}.\footnote{Unless otherwise specified, in the following analysis, $r_i^{NS}$ is always approximated by its lower-bound $r_i^{NS\_l}$ in the unordered SIC case.} Fig. \ref{pA_mu} illustrates how the steady-state point $p_A$ varies with the SINR threshold $\mu$ for the ordered SIC and unordered SIC cases. The result of the capture model \cite{Dai_capture} is also presented for the sake of comparison. Similar to the conditional probability of successful transmission illustrated in Fig. \ref{ri_mu}, we can see from Fig. \ref{pA_mu} that the gains achieved by SIC receivers over the capture model are significant only when $\mu$ is small.

\begin{figure*}[!tp]
\centering
\subfloat[]{
\label{ri_i_0dB}
\includegraphics[width=3.5in,height=2.5in]{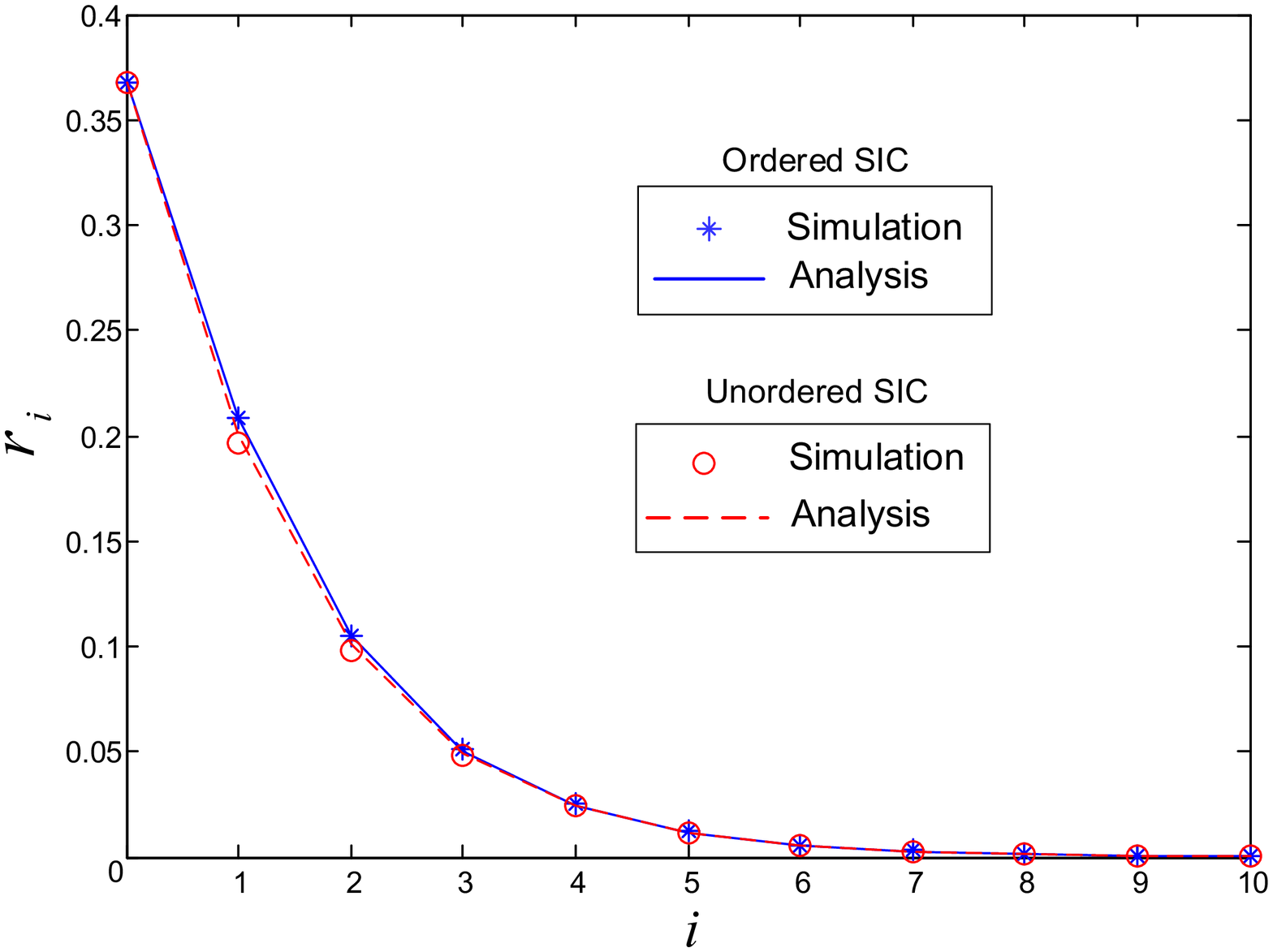}}
\hfil
\subfloat[]{
\label{ri_i_20dB}
\includegraphics[width=3.5in,height=2.5in]{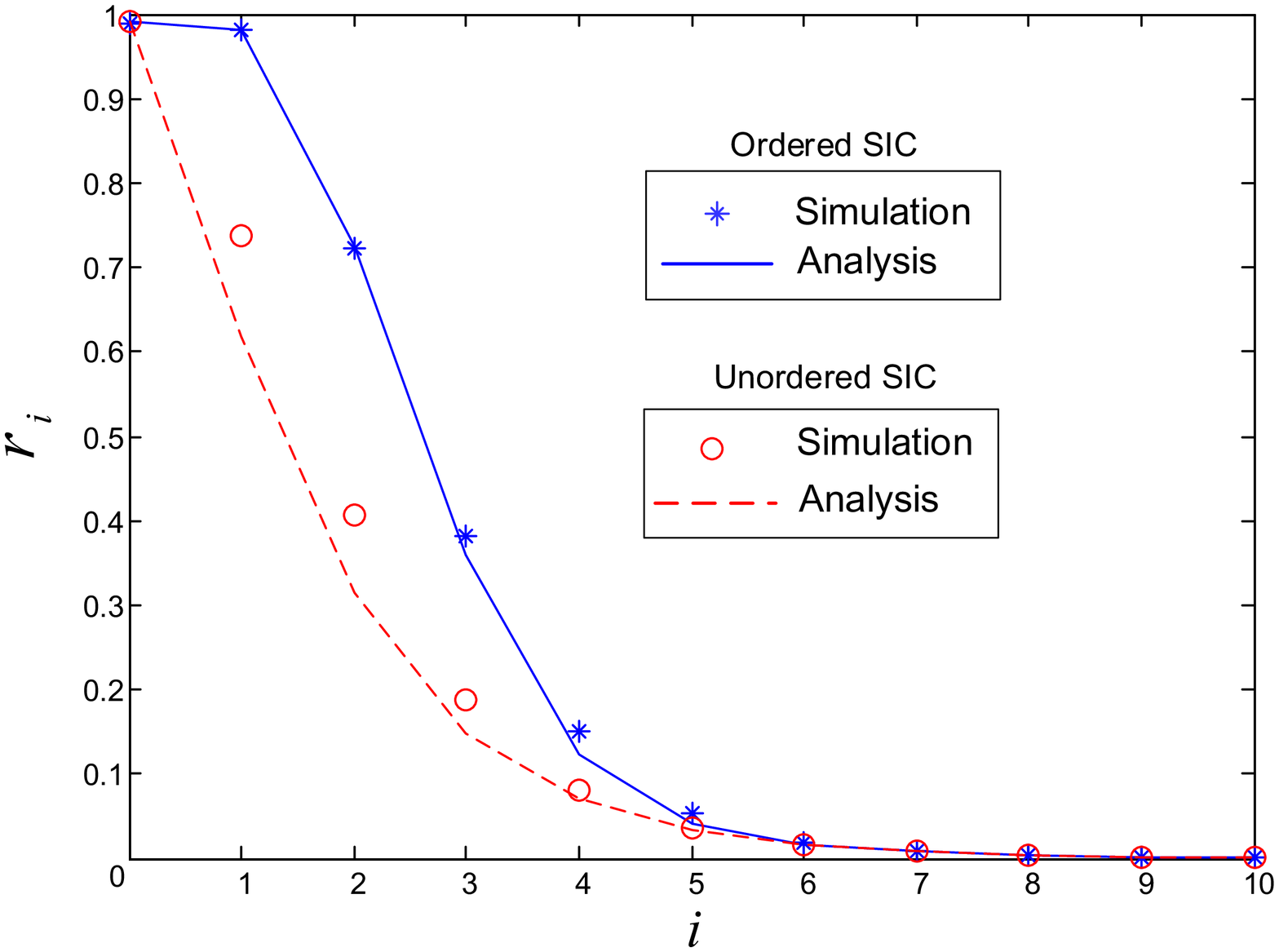}}
\caption{Conditional probability of successful transmission $r_i$ versus number of concurrent packet transmissions $i$. $\mu=1$. (a) $\rho=0$dB. (b) $\rho=20$dB.}
\label{ri_i}
\end{figure*}

\begin{figure*}[!tp]
\centering
\subfloat[]{
\label{pA_q_0dB}
\includegraphics[width=3.5in,height=2.5in]{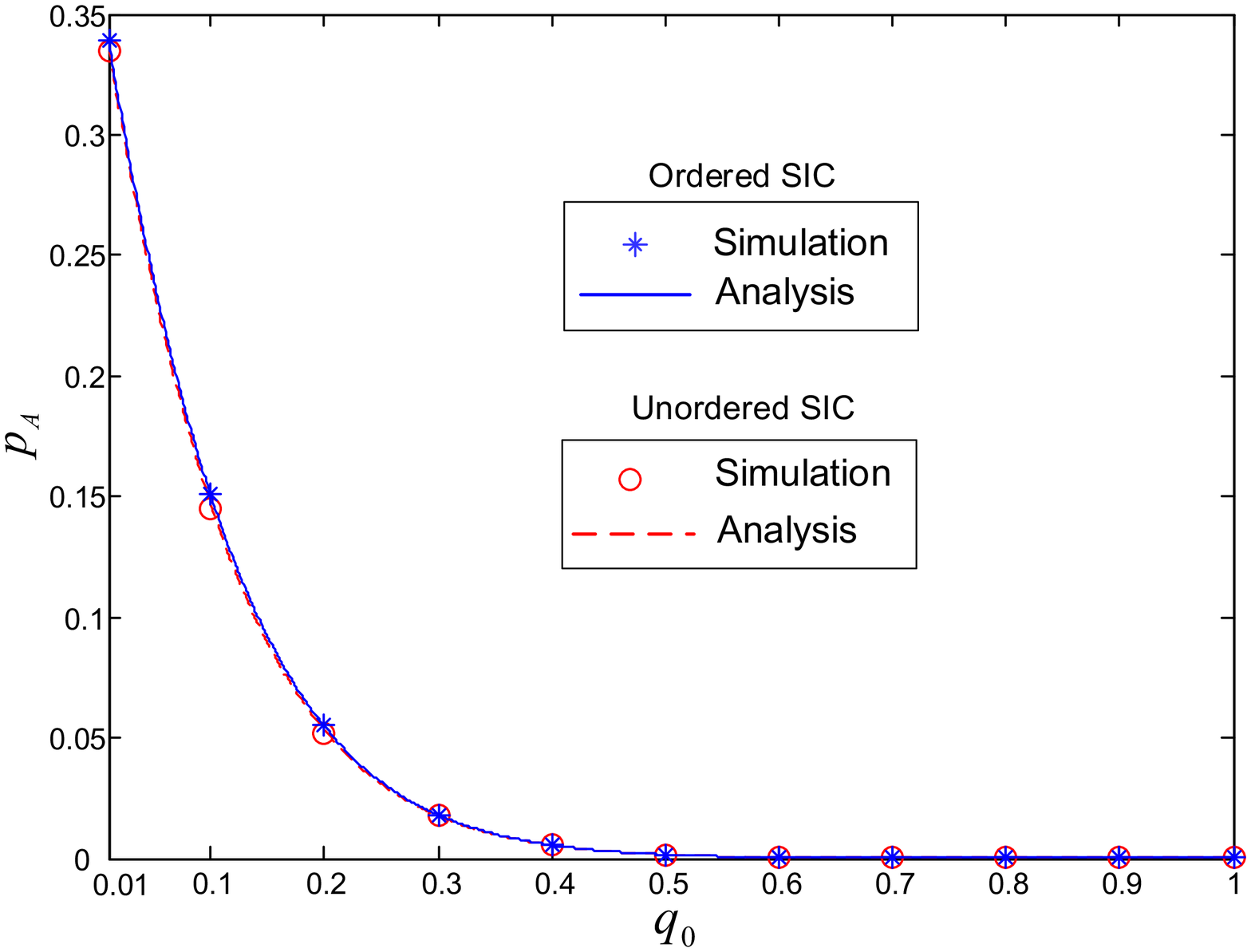}}
\hfil
\subfloat[]{
\label{pA_q_20dB}
\includegraphics[width=3.5in,height=2.5in]{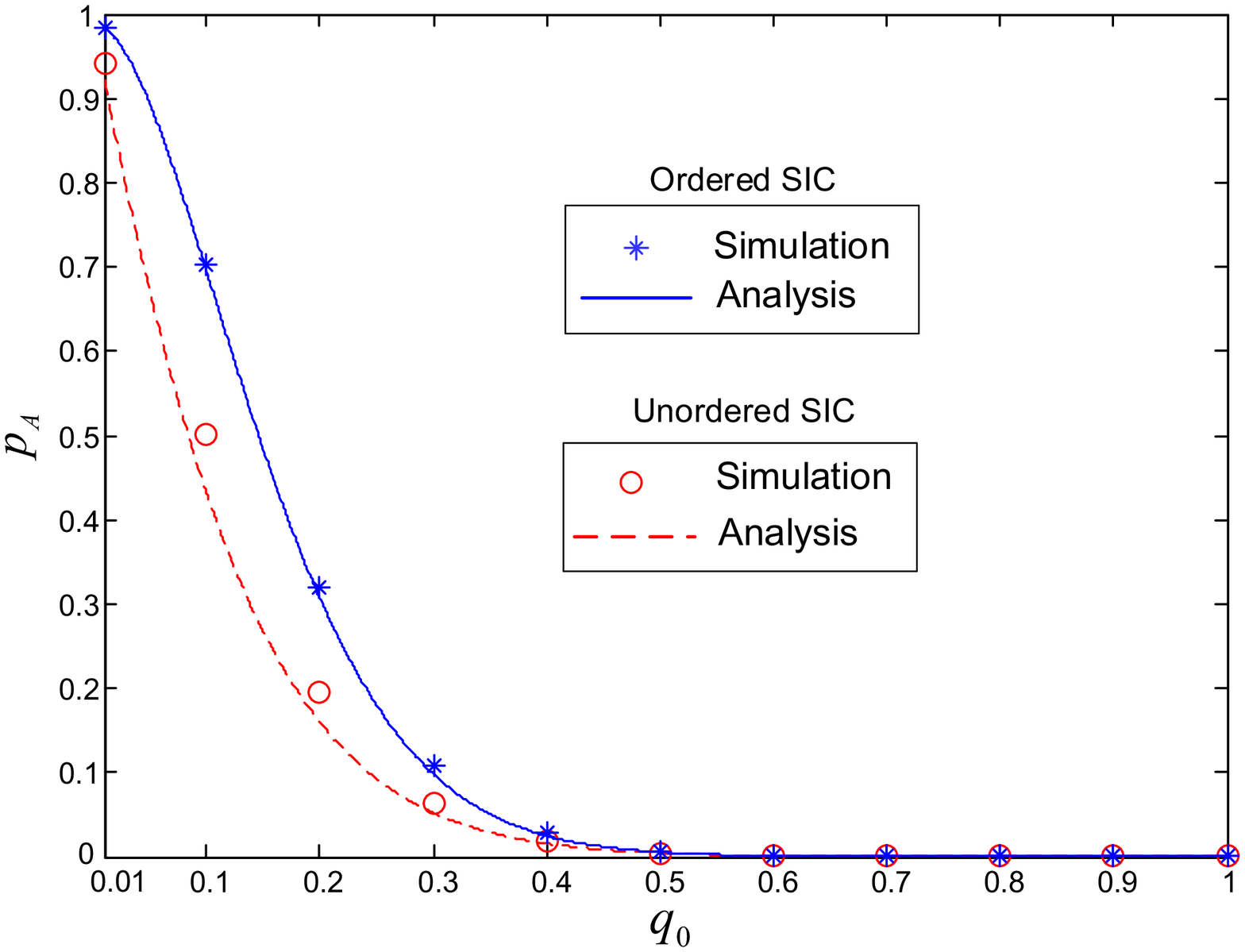}}
\caption{Steady-state point $p_A$ versus transmission probability $q_0$. $n=20$ and $\mu=1$. (a) $\rho=0$dB. (b) $\rho=20$dB.}
\label{pA_q}
\end{figure*}

\subsection{Simulation Results}
In this section, simulation results are presented to verify the preceding analysis. In particular, we consider a saturated slotted Aloha network where each node has the transmission probability $q_0$. The simulation setting is the same as the system model and thus we omit the details here.

In Section \ref{Section3-1}, the conditional probability of successful transmission $r_i$ for ordered SIC has been derived in \eqref{rij_so} and verified by simulation results presented in Fig. \ref{ri_i}. For the unordered SIC case, Fig. \ref{ri_i} shows that the lower-bound of $r_i$ developed in \eqref{rij_sn} is quit close. In Section \ref{Section3-4}, the steady-state point $p_A$ is further derived in \eqref{pA_K0} as the non-zero root of the fixed-point equation of the steady-state probability of successful transmission of HOL packets $p$, which is verified by simulation results presented in Fig. \ref{pA_q}. It can be clearly observed from Fig. \ref{pA_q} that the steady-state point $p_A$ closely depends on the transmission probability of nodes $q_0$. In the next section, we will demonstrate how to optimally choose the transmission probability of nodes $q_0$ to maximize the network throughput and the sum rate.

\section{Maximum Sum Rate}\label{Section4}
It has been shown in Section \ref{Section2-4} that the sum rate is determined by the SINR threshold $\mu$ and the network throughput $\hat{\lambda}_{out}$. In the following subsection, let us first derive the maximum network throughput $\hat{\lambda}_{\max}$ by optimizing the transmission probability of nodes $q_0$.

\subsection{Maximum Network Throughput}\label{Section4-1}
In saturated conditions, the network throughput $\hat{\lambda}_{out}=n \pi_T$. By combining \eqref{pi0} and \eqref{pA_K0}, the network throughput $\hat{\lambda}_{out}$ in saturated conditions can be obtained as
\begin{equation}\label{throughput_K0}
\hat{\lambda}_{out}=n q_0 p_A=n\sum_{i=0}^{n-1}\binom{n-1}{i} r_i \left(1-q_0\right)^{n-1-i} q_0^{i+1},
\end{equation}
which closely depends on the transmission probability of nodes $q_0$. For the maximum network throughput $\hat{\lambda}_{\max}=\max_{0<q_0\leq 1}\hat{\lambda}_{out}$, Appendix \ref{Proof_maxThroughput} shows that in both the ordered SIC and unordered SIC cases, we have
\begin{equation}\label{maxThroughput}
\hat{\lambda}_{\max}=\begin{cases}
n\sum_{i=0}^{n-1}\binom{n-1}{i} r_i \left(1-\hat{q}_0\right)^{n-1-i} \hat{q}_0^{i+1}   & \text{if\;\;}\mu\geq\mu_0 \\
n r_{n-1} & \text{otherwise},
\end{cases}
\end{equation}
which is achieved at
\begin{equation}\label{optimal_q}
q_0^*=\begin{cases}
\hat{q}_0   & \text{if\;\;}\mu\geq\mu_0 \\
1 & \text{otherwise},
\end{cases}
\end{equation}
where $\hat{q}_0$ is the root of
\begin{equation}\label{qm_SO}
n\sum_{i=0}^{n-1}\binom{n-1}{i} r_i \left(1-q_0\right)^{n-2-i} q_0^{i}(1+i-nq_0)=0,
\end{equation}
and $\mu_0$ is the root of
\begin{equation}\label{mu_0}
\dfrac{r_{n-2}}{r_{n-1}}=\dfrac{n}{n-1}.
\end{equation}

\begin{figure*}[!tp]
\centering
\includegraphics[width=4.2in,height=2.5in]{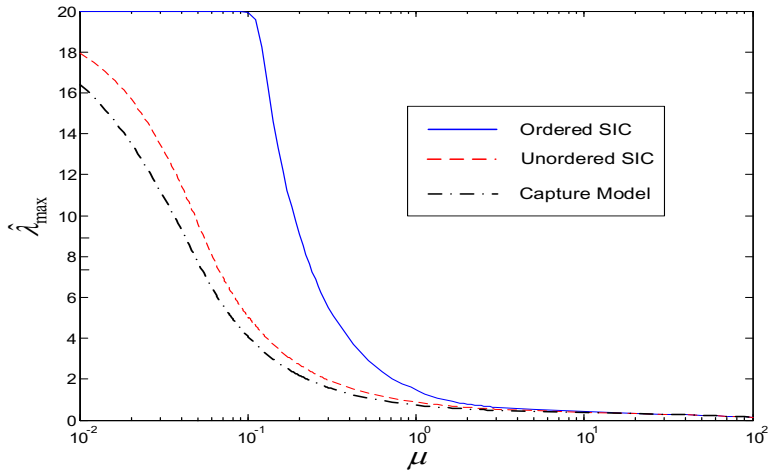}
\caption{Maximum network throughput $\hat{\lambda}_{\max}$ versus SINR threshold $\mu$. $n=20$ and $\rho=20$dB.}
\label{maxThroughput_versus_Mu}
\end{figure*}

We can see from \eqref{maxThroughput} that both $\hat{\lambda}_{\max}^{OS}$ and $\hat{\lambda}_{\max}^{NS}$ depend on the SINR threshold $\mu$. Fig. \ref{maxThroughput_versus_Mu} illustrates how they vary with $\mu$. Similar to the conditional probability of successful transmission $r_{n-1}$ and the steady-state point in saturated conditions $p_A$ shown in Fig. \ref{ri_mu} and Fig. \ref{pA_mu}, significant gains in maximum network throughput can be achieved by SIC receivers over the capture model when $\mu$ is small. Recall that with the capture model \cite{Dai_capture}, we have $\hat{\lambda}_{\max}^{C}\Large|_{\mu=\frac{1}{n}}\stackrel{\text{for large } n}{\approx}n e^{-1}$. It can be further obtained from \eqref{maxThroughput} and \eqref{Corollary_smallthreshold} that
\begin{equation}\label{Corollary_smallthreshold_maxThroughput_SN}
\hat{\lambda}_{\max}^{OS}|_{\mu=\frac{1}{n}}\stackrel{\text{for large } n}{\approx} n, \;\;\;\;\; \hat{\lambda}_{\max}^{NS}|_{\mu=\frac{1}{n}}\stackrel{\text{for large } n}{\approx} n\left(e^{e^{-1}}-1\right),
\end{equation}
both of which are much higher than $n e^{-1}$. With a large $\mu$, nevertheless, the maximum network throughput in both the ordered SIC and unordered SIC cases converges to that of the capture model. As $\mu\rightarrow\infty$, it can be obtained from \eqref{maxThroughput} and \eqref{Corollary_largethreshold_ri_unordered} that
$\lim_{\mu\rightarrow\infty}\tfrac{\hat{\lambda}_{\max}^{OS}}{\hat{\lambda}_{\max}^C}=\lim_{\mu\rightarrow\infty}\tfrac{\hat{\lambda}_{\max}^{NS}}{\hat{\lambda}_{\max}^C}=1$.

It can be clearly seen from Fig. \ref{maxThroughput_versus_Mu} that similar to the capture model, with SIC receivers, the maximum network throughput $\hat{\lambda}_{\max}$ also increases as the SINR threshold $\mu$ decreases. Such an improvement on maximum network throughput is, nevertheless, achieved at the cost of a smaller information encoding rate that can be supported for reliable communications, i.e., $R=\log_2(1+\mu)$. In the next subsection, we will further study how to maximize the sum rate by properly choosing the SINR threshold $\mu$.

\begin{figure*}[!tp]
\centering
\includegraphics[width=4.2in,height=2.5in]{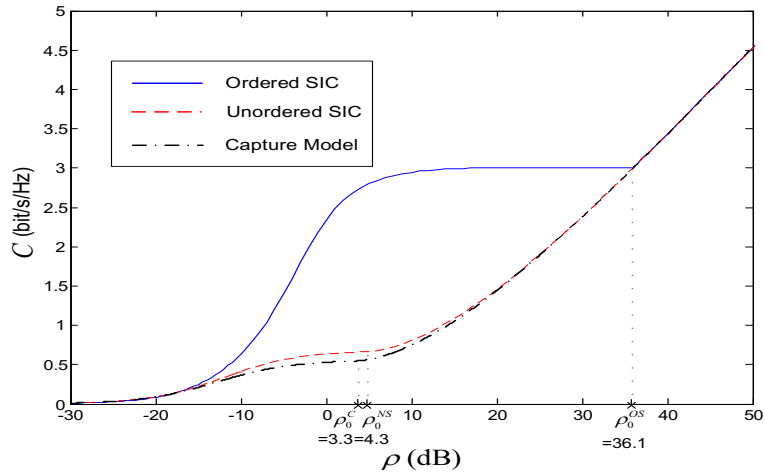}
\caption{Maximum sum rate $C$ versus mean received SNR $\rho$. $n=20$.}
\label{C_all_models_1}
\end{figure*}

\subsection{Maximum Sum Rate}\label{Section4-2}
By combining \eqref{maxRate_1} and \eqref{maxThroughput}, the maximum sum rate can be further written as $C=\max\left(C_1,C_2\right)$, where $C_1=\max_{\mu\geq\mu_0} f_1(\mu)$ and $C_2=\max_{\mu<\mu_0} f_2(\mu)$, in which
\begin{equation}\label{f_1_expre}
f_1(\mu)=\hat{\lambda}_{\max}^{\mu\geq\mu_0}\log_2(1+\mu)=n\sum_{i=0}^{n-1}\binom{n-1}{i} r_i \left(1-\hat{q}_0\right)^{n-1-i} \hat{q}_0^{i+1}\cdot \log_2(1+\mu)
\end{equation}
for $\mu\in[\mu_0,\infty)$ and
\begin{equation}\label{f_2_expre}
f_2(\mu)=\hat{\lambda}_{\max}^{\mu<\mu_0}\log_2(1+\mu)=n r_{n-1}\cdot \log_2(1+\mu)
\end{equation}
for $\mu\in(0,\mu_0)$. Denote $\mu_h=\arg\max f_1(\mu)$ and $\mu_l=\arg\max f_2(\mu)$. Note that $\mu_h$ and $\mu_l$ can be numerically obtained given the expressions of $f_1(\mu)$ and $f_2(\mu)$. Appendix \ref{Derivation of C} shows that for both the ordered SIC and unordered SIC cases, the maximum sum rate is given by
\begin{equation}\label{C_expression}
C=\begin{cases}
f_1(\mu_h)   & \text{if\;\;}\rho\geq\rho_0 \\
f_2(\mu_l) & \text{otherwise},
\end{cases}
\end{equation}
which is achieved when the SINR threshold $\mu$ is set as
\begin{equation}\label{optimal_threshold_expression}
\mu^{*}=\begin{cases}
\mu_h   & \text{if\;\;}\rho\geq\rho_0 \\
\mu_l & \text{otherwise},
\end{cases}
\end{equation}
where $\rho_0$ is the root of $f_1(\mu_h)=f_2(\mu_l)$.

The maximum sum rate $C$ for both the ordered SIC and unordered SIC cases is illustrated in Fig. \ref{C_all_models_1}. For the sake of comparison, the result of the capture model \cite{Dai_capture} is also presented. In contrast to Fig. \ref{maxThroughput_versus_Mu} where the maximum network throughput of the unordered SIC is shown to be significantly higher than that of the capture model, we can see from Fig. \ref{C_all_models_1} that only marginal gains in the maximum sum rate can be achieved by the unordered SIC over the capture model. For the ordered SIC case, substantial gains can still be observed at moderate values of the mean received SNR $\rho$, i.e., $-10$dB$<\rho<35$dB. The gap, nevertheless, diminishes at the high SNR region, i.e., $\rho\geq \rho_0^{OS}$, where the maximum sum rates of all the three receivers have the high-SNR slope of $e^{-1}$ as $\rho\rightarrow\infty$.

\begin{figure*}[!tp]
\centering
\subfloat[]{
\label{C_all_models_2}
\includegraphics[width=3.5in,height=2.5in]{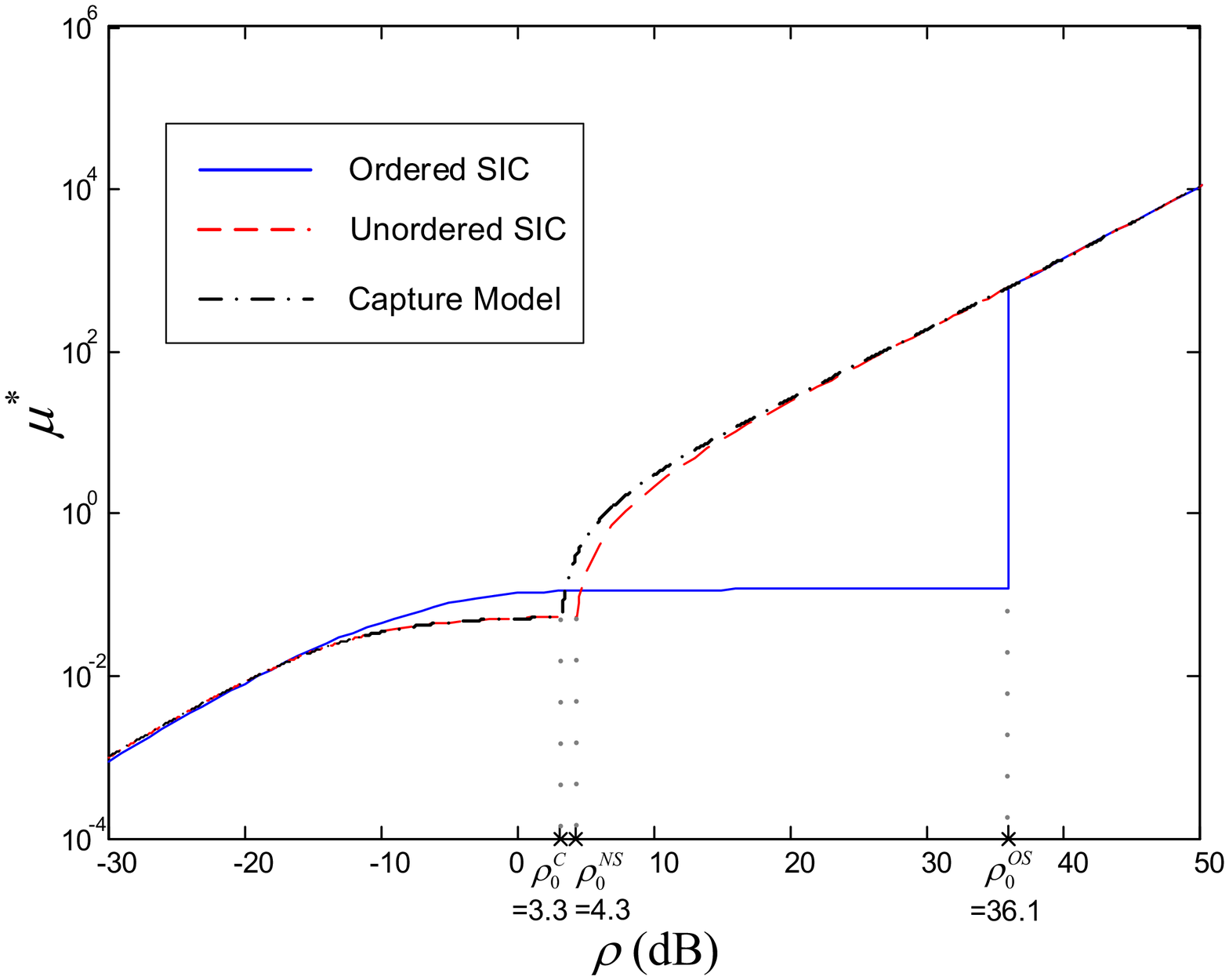}}
\hfil
\subfloat[]{
\label{C_all_models_3}
\includegraphics[width=3.5in,height=2.5in]{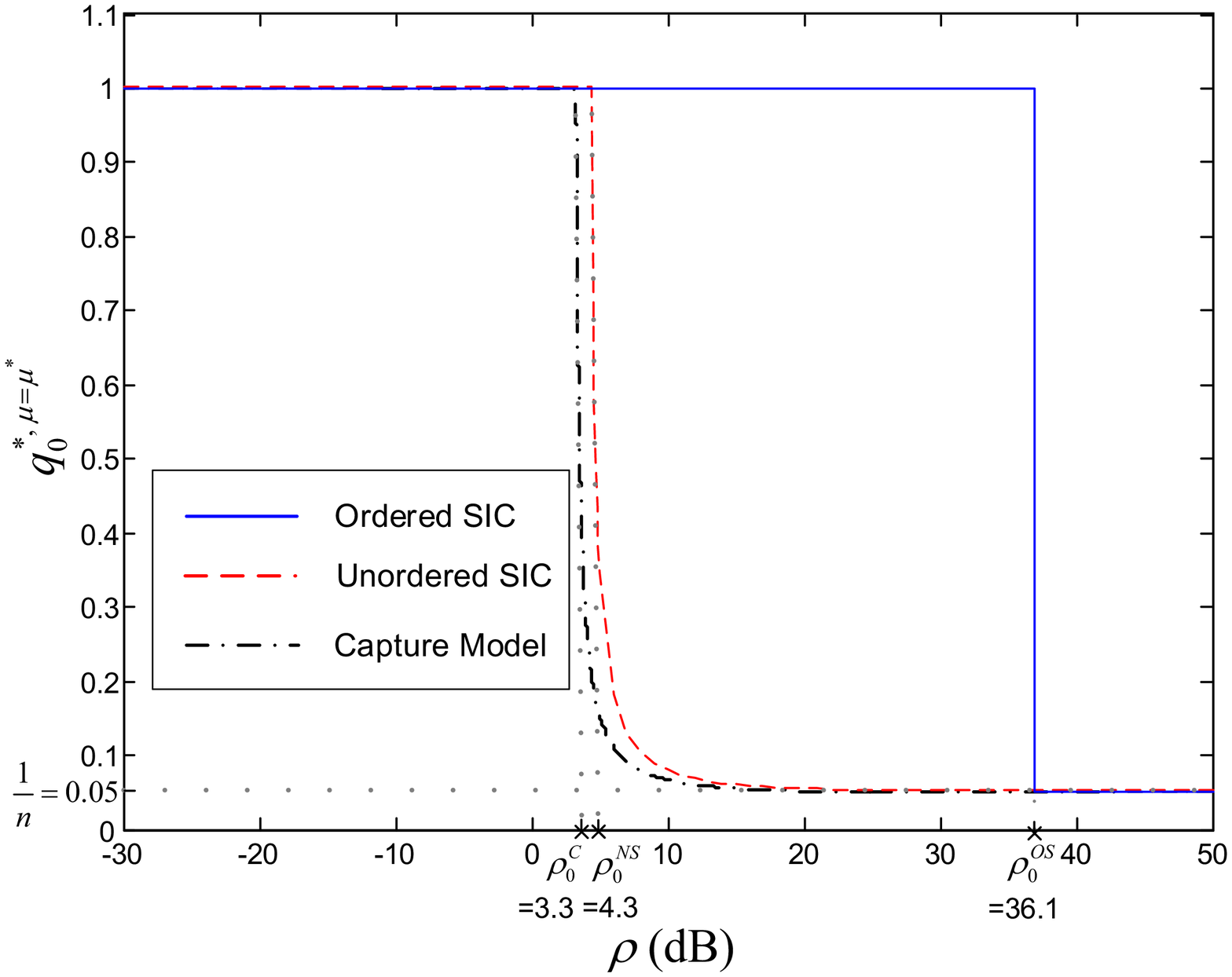}}
\caption{(a) Optimal SINR threshold $\mu^*$ and (b) optimal transmission probability $q_0^{*,\mu=\mu^*}$. $n=20$.}
\label{C_all_models}
\end{figure*}

It is also interesting to note from Fig. \ref{C_all_models_1} that ordering is crucial for the rate performance of slotted Aloha networks with SIC receivers. Without proper ordering of packets, the maximum sum rate could be drastically degraded and becomes comparable to that of the capture model.
To achieve the maximum sum rate, the SINR threshold $\mu$ needs to be carefully tuned according to the mean received SNR $\rho$ based on \eqref{optimal_threshold_expression}. Fig. \ref{C_all_models_2} illustrates the optimal SINR threshold $\mu^*$ for the ordered SIC, unordered SIC and the capture model cases. It can be clearly seen from Fig. \ref{C_all_models_2} that in all three cases, the SINR threshold $\mu$, or equivalently, the information encoding rate $R$, should be properly enlarged as the mean received SNR $\rho$ increases. Moreover, by combining \eqref{optimal_q} and \eqref{optimal_threshold_expression}, we can further obtain the optimal transmission probability $q_0^{*,\mu=\mu^*}$ for maximizing the sum rate as a function of the mean received SNR $\rho$. As Fig. \ref{C_all_models_3} illustrates, at the low SNR region, the optimal transmission probability is $1$, indicating that all the nodes should persistently transmit their packets. As the mean received SNR $\rho$ increases, nevertheless, the transmission probability of each node should be reduced accordingly, and converges to $\frac{1}{n}$ as $\rho\rightarrow \infty$.\footnote{Note that in practice, the adaptive update of the transmission probability $q_0$ and the information encoding rate $R$ of each node can be performed through the feedback from the common receiver. In cellular systems, for instance, as each node associates with the base station (BS) upon joining the network, the BS can count the number of nodes, calculate the optimal parameter setting and broadcast it periodically. Each node can then update its parameters accordingly. Such a feedback-based update process can also be implemented in IEEE 802.11 networks where the access point serves as the common receiver in each basic service set.}

\subsection{Simulation Results}

\begin{figure*}[!tp]
\centering
\subfloat[]{
\label{throughput_smallMu}
\includegraphics[width=3.5in,height=2.5in]{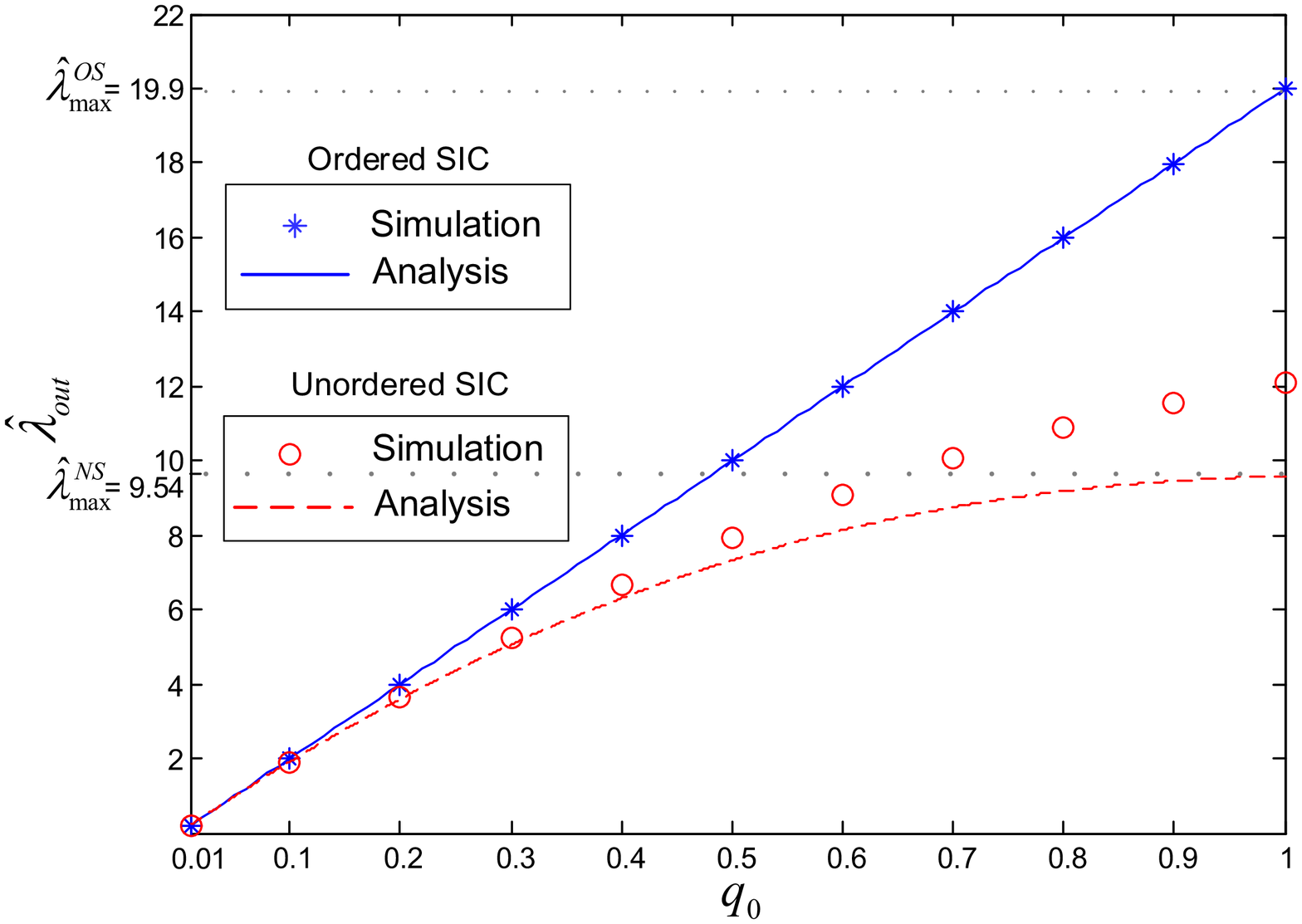}}
\hfil
\subfloat[]{
\label{throughput_Mu=1}
\includegraphics[width=3.5in,height=2.5in]{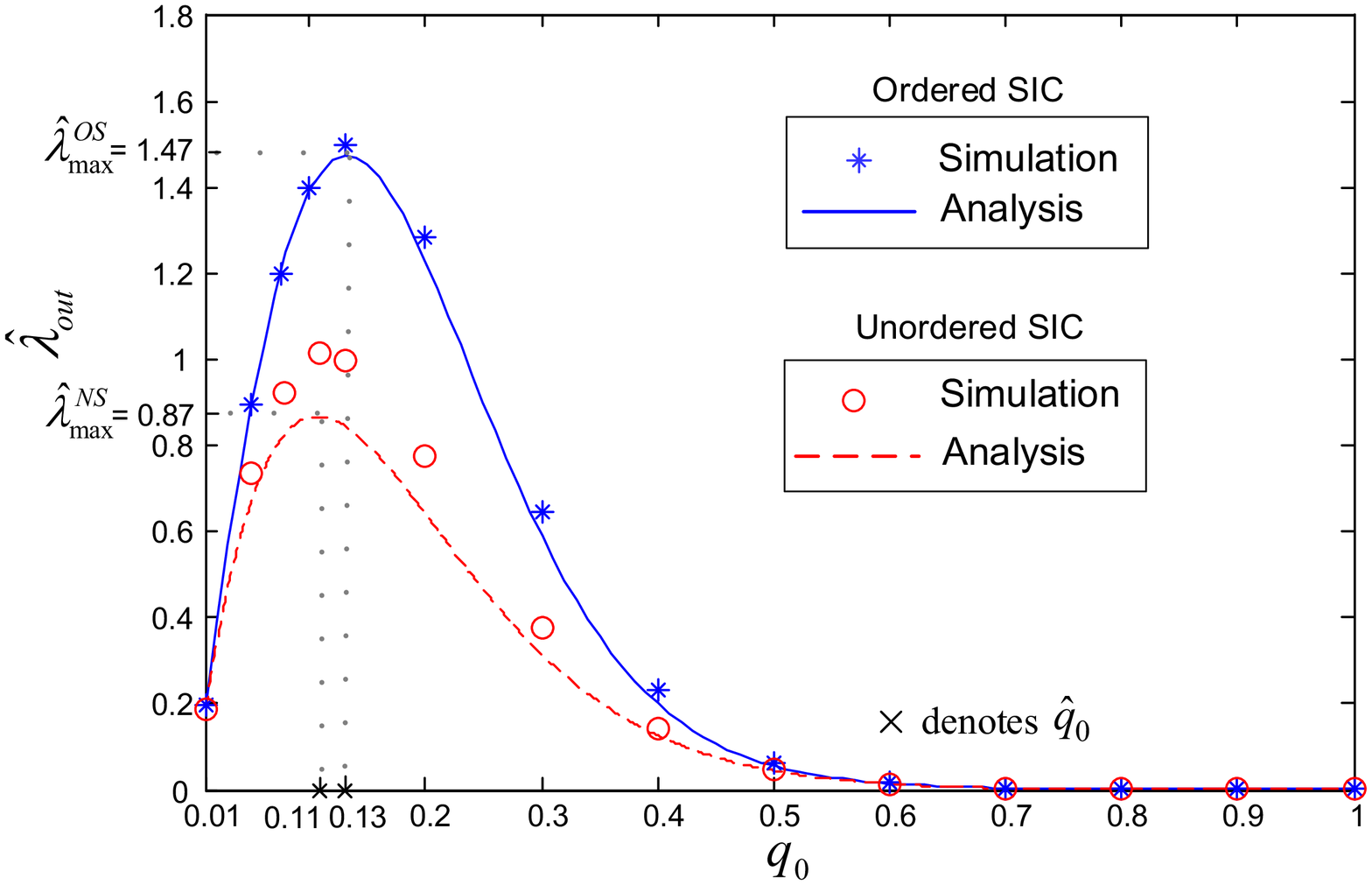}}
\caption{Network throughput $\hat{\lambda}_{out}$ versus transmission probability $q_0$. $n=20$ and $\rho=20$dB. (a) $\mu=0.05$. (b) $\mu=1$.}
\label{throughput_simulation}
\end{figure*}

In this section, simulation results are presented to verify the preceding analysis. Fig. \ref{throughput_simulation} illustrates the network throughput performance. The network throughput $\hat{\lambda}_{out}$ has been derived as a function of the transmission probability of nodes $q_0$ in \eqref{throughput_K0}. According to \eqref{maxThroughput}, if the SINR threshold $\mu\geq\mu_0$, the maximum network throughput $\hat{\lambda}_{\max}$ is achieved when $q_0$ is set to be $\hat{q}_0$. Otherwise, $\hat{\lambda}_{\max}$ is achieved when $q_0=1$. With $n=20$ and $\rho=20$dB, we have $\mu_0^{OS}=0.1205$ and $\mu_0^{NS}=0.0532$ by substituting \eqref{rij_so} and \eqref{rij_sn} into \eqref{mu_0}, respectively, for the ordered SIC and unordered SIC cases. As we can see from Fig. \ref{throughput_smallMu}, with $\mu=0.05<\mu_0$, the network throughput in both cases monotonically increases with $q_0$, and is maximized at $q_0=1$. With $\mu=1>\mu_0$, on the other hand, the network throughput is maximized at $q_0=\hat{q}_0$ which is given in \eqref{qm_SO}, as Fig. \ref{throughput_Mu=1} illustrates. Simulation results presented in Fig. \ref{throughput_simulation} verify that the analysis serves as a good lower-bound in the unordered SIC case, and is accurate with ordered SIC.

\begin{figure*}[!tp]
\begin{minipage}{.5\linewidth}
\centering
\subfloat[]{\label{Simulation_Sum_Rate_1}\includegraphics[width=3.5in,height=2.5in]{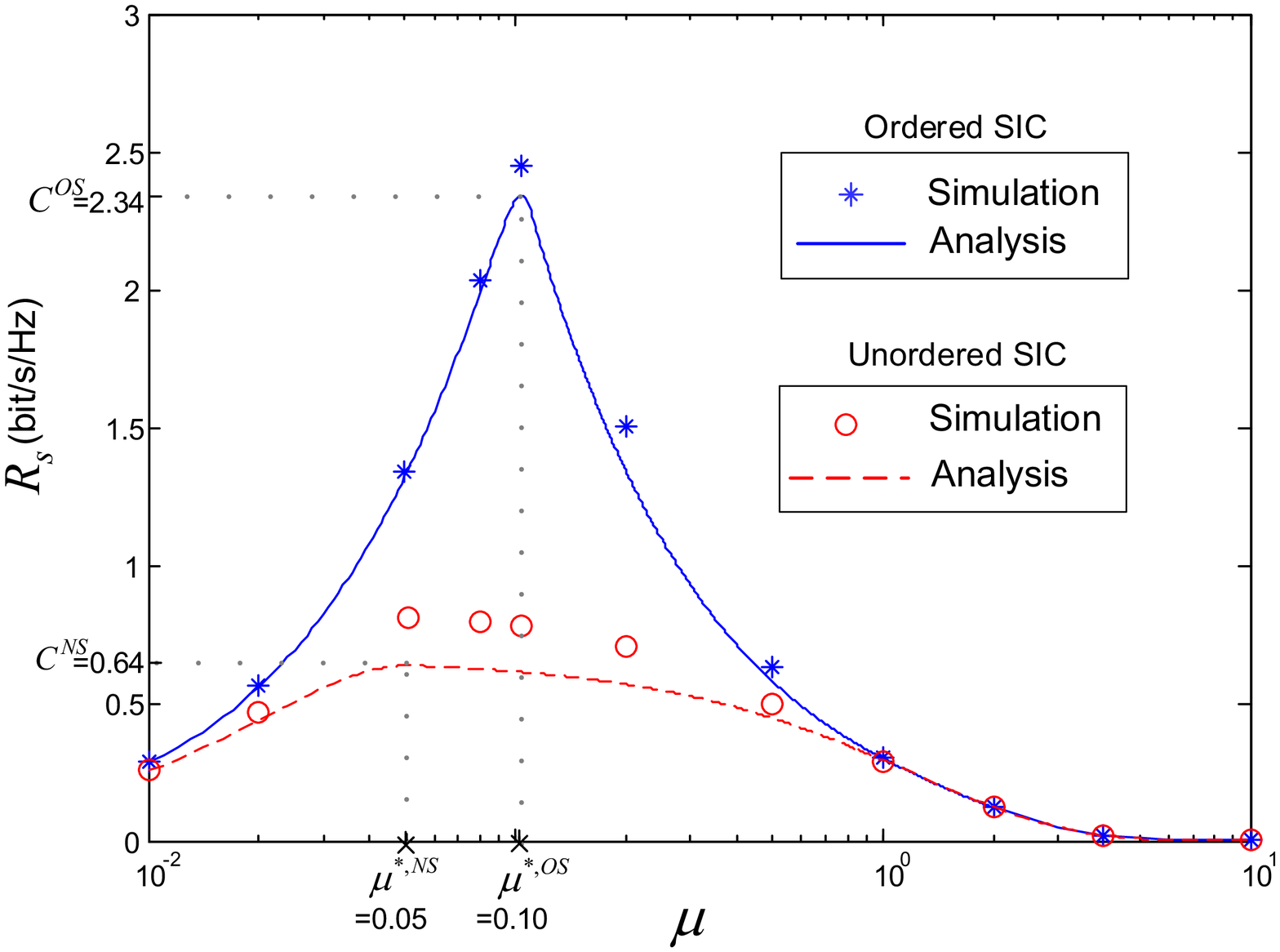}}
\end{minipage}%
\begin{minipage}{.5\linewidth}
\centering
\subfloat[]{\label{Simulation_Sum_Rate_2}\includegraphics[width=3.5in,height=2.5in]{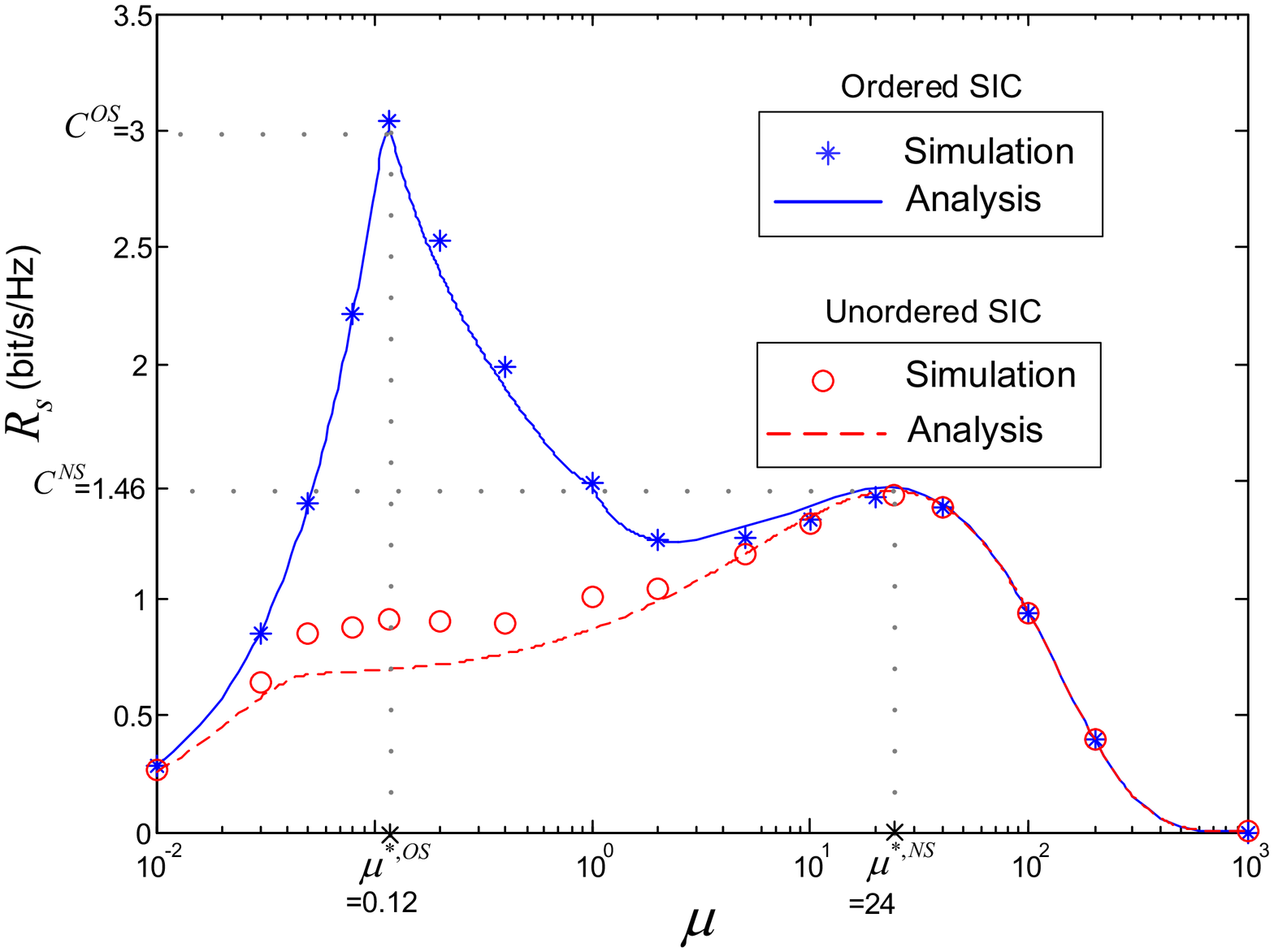}}
\end{minipage}\par\medskip
\centering
\subfloat[]{\label{Simulation_Sum_Rate_3}\includegraphics[width=3.5in,height=2.5in]{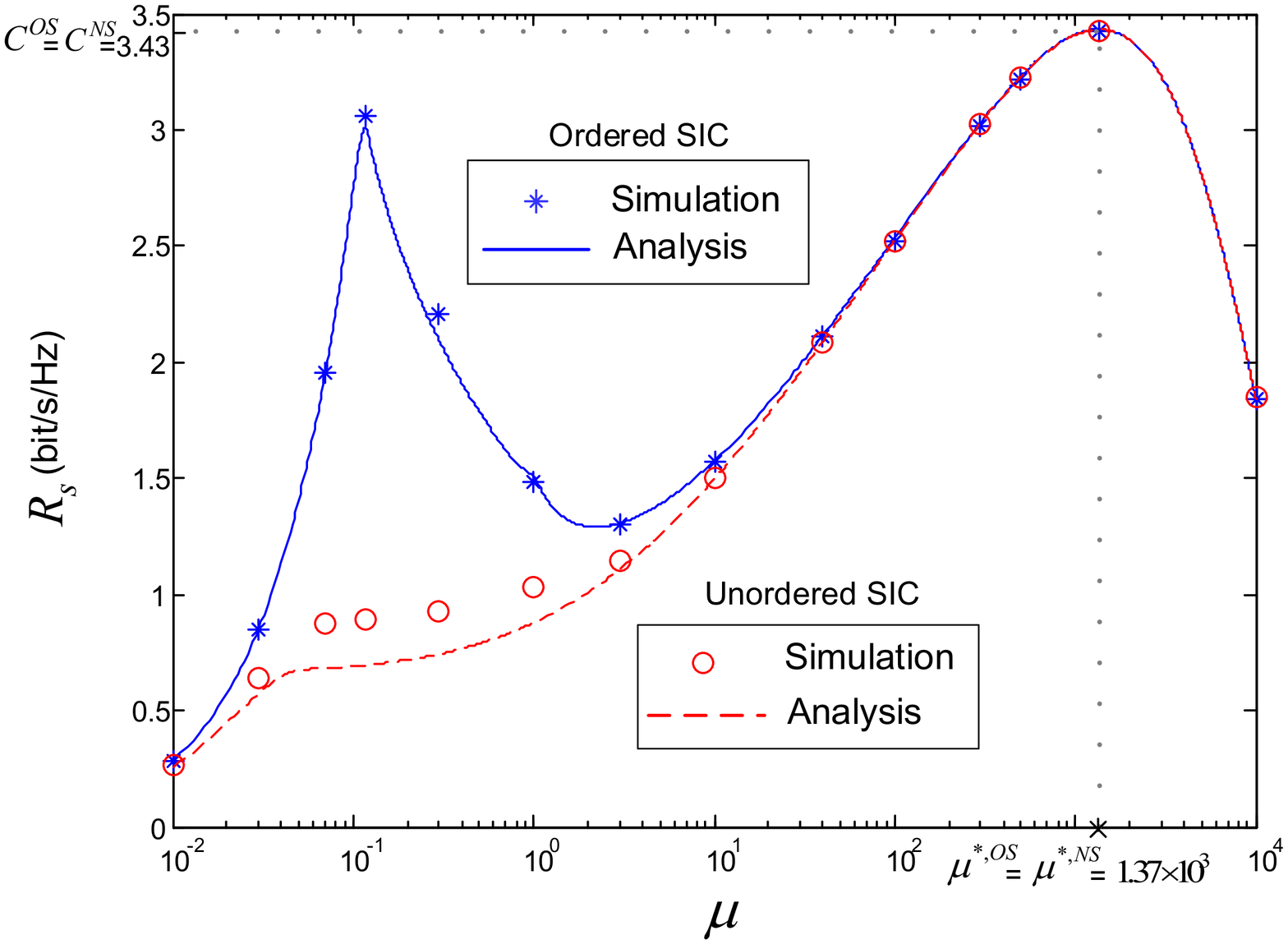}}

\caption{Sum rate $R_s$ versus SINR threshold $\mu$. $n=20$ and $q_0=q_0^*$. (a) $\rho=0$dB. (b) $\rho=20$dB. (c) $\rho=40$dB.}
\label{Simulation_Sum_Rate}
\end{figure*}

It is further shown in Section \ref{Section4-2} that as both the information encoding rate $R$ and the maximum network throughput $\hat{\lambda}_{\max}$ depend on the SINR threshold $\mu$, the sum rate can be maximized by optimizing the SINR threshold $\mu$. As Fig. \ref{Simulation_Sum_Rate} illustrates, the sum rate performance is sensitive to the setting of $\mu$ in both ordered SIC and unordered SIC cases. To achieve the maximum sum rate, the SINR threshold $\mu$ should be properly set according to the mean received SNR $\rho$. The expressions of the maximum sum rate $C$ and the corresponding optimal SINR threshold $\mu^*$  are given in (\ref{C_expression})-(\ref{optimal_threshold_expression}), and verified by simulation results presented in Fig. \ref{Simulation_Sum_Rate}.

\section{Discussions}\label{Section5}
So far we have derived the maximum sum rates of slotted Aloha networks with SIC receivers, and the corresponding optimal setting including the optimal SINR threshold and the optimal transmission probability of nodes. In this section, we will further demonstrate the effect of MPR on the sum rate performance of slotted Aloha networks, and compare it to the ergodic sum capacity of multiple access fading channels.

\subsection{Effect of MPR}\label{Section5-1}
Note that both the SIC receivers and the capture model have the so-called MPR capability: If the SINR threshold $\mu$ is small enough, then multiple packets can be successfully decoded at each time slot. As a result, the network throughput performance is expected to be substantially enhanced compared to the classical collision model where at most one packet can be successfully decoded at each time slot.

To further see the effect of MPR on the sum rate performance, let us consider the classical collision model where a packet transmission is successful only if there are no concurrent transmissions. Appendix \ref{Aloha_collision-model} shows that in this case, the maximum sum rate is given by
\begin{equation}\label{maximum sum rate_collision}
C^{collision}= \exp\left (-1-\tfrac{e^{{\mathbb W}_{0}(\rho)}-1}{\rho}\right)\cdot \log_2(e^{{\mathbb W}_{0}(\rho)}),
\end{equation}
which is achieved when the SNR threshold is set to be $\mu^{*,collision}=e^{{\mathbb W}_{0}(\rho)}-1$. The corresponding maximum network throughput is given by
\begin{equation}\label{maximum throughput_collision}
\hat{\lambda}_{\max}^{\mu=\mu^*,collision}=\exp\left(-1-\tfrac{e^{{\mathbb W}_{0}(\rho)}-1}{\rho}\right).
\end{equation}

\begin{figure*}[!tp]
\centering
\subfloat[]{
\label{maxThroughput_comparison}
\includegraphics[width=3.5in,height=2.5in]{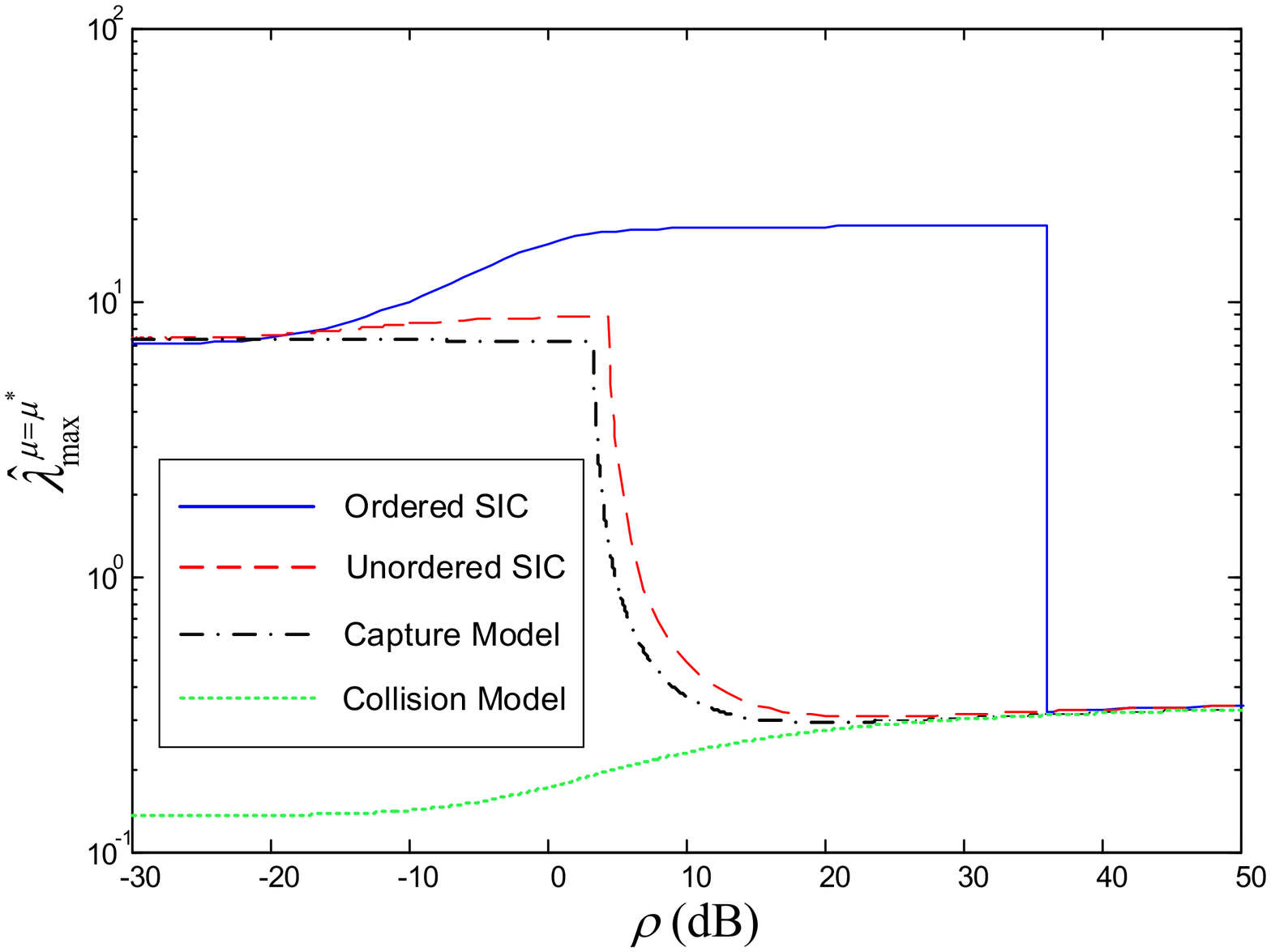}}
\hfil
\subfloat[]{
\label{maxR_comparison}
\includegraphics[width=3.5in,height=2.5in]{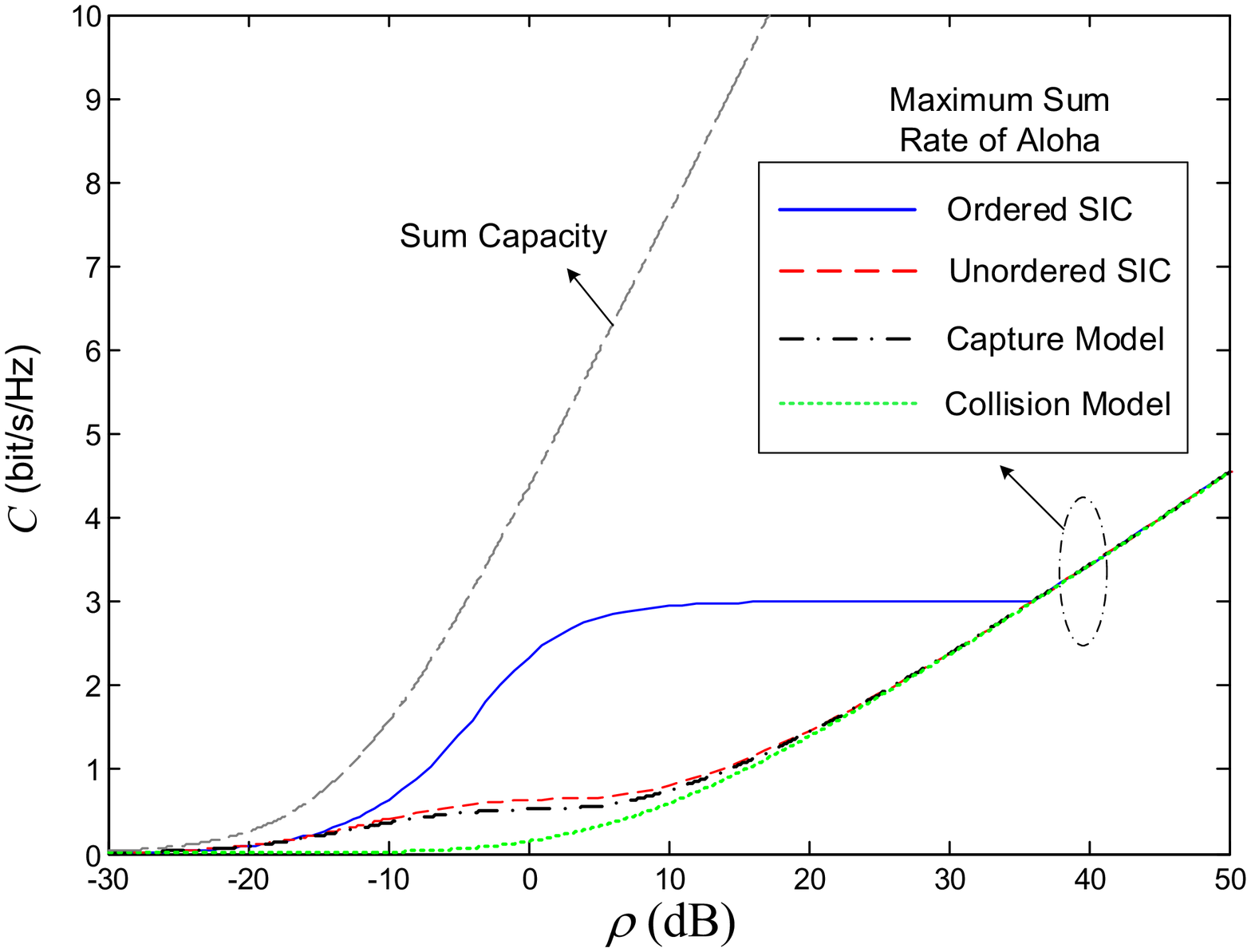}}
\caption{Network performance of slotted Aloha with collision model, capture model and SIC receivers. $n=20$. (a) Maximum network throughput $\hat{\lambda}_{\max}^{\mu=\mu^*}$ and (b) maximum sum rate $C$ versus mean received SNR $\rho$.}
\label{channel_comparison}
\end{figure*}

Fig. \ref{channel_comparison} illustrates the maximum sum rate and the corresponding maximum network throughput with the collision model. For the sake of comparison, the results of the capture model and SIC receivers are also presented. As we can see from Fig. \ref{maxThroughput_comparison}, at the low SNR region, the maximum network throughput of the collision model is significantly lower than that of SIC receivers and the capture model, which can be much higher than $1$ thanks to MPR. The gap, nevertheless, diminishes as the mean received SNR $\rho$ increases. All of them approach $e^{-1}$ as $\rho\to\infty$.

Similarly, Fig. \ref{maxR_comparison} shows that the maximum sum rates of all four receivers have the same high-SNR slope of $e^{-1}$, and the rate difference becomes negligible when the mean received SNR $\rho$ is large. At the low SNR region, in contrast to the significant throughput improvement, only marginal gains in the maximum sum rate can be achieved by the SIC receivers and the capture model. The rate difference between the collision model and the capture model is below $0.5$ bit/s/Hz for the whole SNR region, which suggests that despite an overly pessimistic estimation on the network throughput, the collision model serves as a good approximation for the capture model when analyzing the sum rate performance of Aloha networks.

Recall that it has been shown in Fig. \ref{C_all_models_2} that at the low SNR region, the optimal SINR thresholds with the SIC receivers and the capture model are small, indicating that each packet has a small encoding rate. Therefore, although much more packets can be successfully decoded compared to the collision model, the gain in the maximum sum rate is limited. At the high SNR region, on the other hand, both the SIC receivers and the capture model reduce to the collision model as the optimal SINR thresholds are much higher than $1$. We can conclude from Fig. \ref{channel_comparison} that although the network throughput performance at the low SNR region can always be significantly improved thanks to MPR, the rate gain could be marginal, and becomes negligible when the mean received SNR is large.

\subsection{Rate Loss Due to Random Access}
For an $n$-user multiple access fading channel, it has been well known that the ergodic sum capacity without the channel state information at the transmitter side (CSIT) is given by $C_{sum}=\mathbb{E}\left[\log_2\left(1+\sum_{k=1}^n |h_k|^2\rho\right)\right]$, where the expectation is taken over all fading states of users. To achieve the ergodic sum capacity, an SIC receiver should be adopted, and each node's encoding rate is determined by its decoding order, with codewords long enough to cover multiple fading states \cite{Tse}. With Aloha, in contrast, as each node transmits with a certain probability, the subset of active nodes is time-varying, and supposed to be unknown at the transmitter side. Therefore, without coordination rate allocation among nodes cannot be performed, and a uniform information encoding rate is assumed for each packet with the packet-based encoding, i.e., each packet contains one codeword lasting for one time slot.

Fig. \ref{maxR_comparison} illustrates the ergodic sum capacity in comparison to the maximum sum rates of slotted Aloha under various receivers. It can be clearly seen from Fig. \ref{maxR_comparison} that even with the capacity-achieving receiver structure, the ordered SIC, the maximum sum rate of slotted Aloha is still significantly lower than the ergodic sum capacity. Specifically, different high-SNR slopes are observed, i.e., $1$ for the ergodic sum capacity and $e^{-1}$ for the maximum sum rates of slotted Aloha. With the packet-based encoding/decoding and a uniform encoding rate of each packet, even if the ordered SIC is adopted, the sum rate performance of slotted Aloha is still bottlenecked by the first decoded packet, which can be improved by increasing the mean received SNR $\rho$ at the high SNR region only when there are no concurrent packet transmissions. It has been shown in Fig. \ref{C_all_models_3} that as $\rho$ increases, the optimal transmission probability of each node converges to $1/n$. As a result, the high-SNR slope of the maximum sum rates of slotted Aloha, which is determined by the probability that there is a single packet transmission in each time slot at the high SNR region, is only $e^{-1}$. Here we can see that the huge rate loss of slotted Aloha at the high SNR region has its root in the contention of packets caused by uncoordinated random transmissions of nodes.



\section{Conclusion}\label{Section6}
In this paper, the maximum sum rates of slotted Aloha with two representative SIC receivers, ordered SIC and unordered SIC, are analyzed by assuming that the received SNRs of nodes' packets are exponentially distributed with the mean received SNR $\rho$. The analysis shows that to achieve the maximum sum rate, the transmission probability of each node and the SINR threshold, or equivalently, the encoding rate of each packet, should be adaptively adjusted according to the mean received SNR $\rho$. The maximum sum rate with ordered SIC is found to be significantly higher than that with unordered SIC only at moderate values of $\rho$. At low and high SNR regions, both of them could become comparable to the capture model.

The comparison to the classical collision model further reveals that although substantial gains in network throughput can be achieved by SIC receivers and the capture model at the low SNR region thanks to MPR, the rate difference could be limited, and they all reduce to the collision model with the high-SNR slope of $e^{-1}$ when the mean received SNR $\rho$ is large. It shows that without proper rate allocation, even with the capacity-achieving receiver structure, the maximum sum rate of slotted Aloha could still be far below the ergodic sum capacity of fading channels, indicating that the rate loss is significant due to uncoordinated random transmissions of nodes.

\appendices
\section{Derivation of \eqref{y_i^l}-\eqref{Mu>1}}\label{Appendix_A}
The conditional probability that the packet at the $l$th iteration can be successfully decoded given that there are $i$ concurrent packet transmissions, $y_i^{(l)}$, can be written as
\begin{equation}\label{y_i^l_de}
y_i^{(l)} =\textrm{Pr}\left\{\frac{|h_{l:i+1}|^2}{\sum_{s=l+1}^{i+1} |h_{s:i+1}|^2+1/\rho}\geq\mu\right\},
\end{equation}
where $|h_{1:i+1}|^2\geq |h_{2:i+1}|^2\geq \cdots\geq |h_{i+1:i+1}|^2$ denotes $\{|h_k|^2\}_{k=1,2,\ldots,i+1}$ in descending order\footnote{Note that the order of received power of packets is solely determined by their small-scale fading gains because the effect of large-scale fading is removed according to \eqref{power_control}.}. \eqref{y_i^l_de} can be further written as
\begin{equation}\label{Appendix_eq3}
y_i^{(l)}=\textrm{Pr}\left\{Z\leq\dfrac{X}{\mu}-\dfrac{1}{\rho}\right\}=\displaystyle\int_0^\infty\displaystyle\int_0^{\frac{x}{\mu}{-}\frac{1}{\rho}} p_{X,Z}(x,z) dzdx,
\end{equation}
where $X=|h_{l:i+1}|^2$ and $Z=\sum_{s=l+1}^{i+1} |h_{s:i+1}|^2$, and the joint PDF $p_{X,Z}(x,z)$ is given by \cite{Yang}
\begin{equation}\label{Appendix_eq1}
p_{X,Z}(x,z)=S\cdot \exp\left(-lx-z\right)\sum_{k=0}^{i+1-l}\binom{i+1-l}{k}(-1)^k (z-kx)^{i-l}U(z-kx),
\end{equation}
for $x>0$ and $0<z<(i+1-l)x$,
with
$S=\tfrac{(i+1)!}{(i+1-l)!(l-1)!(i-l)!}$,
and 
\begin{equation}\label{Appendix_eq6}
U(a)=\begin{cases}
1 & a\geq 0 \\
0 & \text{otherwise}.
\end{cases}
\end{equation}
Otherwise, $p_{X,Z}(x,z)=0$.
For $l=1,...,i+1$, we consider the following two cases:

1) If $l\leq i+1-\tfrac{1}{\mu}$, we have $(i+1-l)x > \tfrac{x}{\mu}-\tfrac{1}{\rho}$. By combining \eqref{Appendix_eq1} and \eqref{Appendix_eq3}, $y_i^{(l)}$ can be written as
\begin{align}\label{Appendix_eq4}
y_i^{(l)}&=S \cdot \displaystyle\int_{\frac{\mu}{\rho}}^\infty\displaystyle\int_0^{\frac{x}{\mu}{-}\frac{1}{\rho}}\sum_{k=0}^{i+1-l}\binom{i+1-l}{k}(-1)^k\exp\left(-lx-z\right)(z-kx)^{i-l}U(z-kx)dzdx\notag\\
&=\dfrac{(i{+}1)!}{(i{+}1{-}l)!(l{-}1)!}\displaystyle \sum_{k=0}^{\left\lceil1/\mu\right\rceil{-}1} \binom{i{+}1{-}l}{k}\dfrac{(-1)^k}{l{+}k}\exp\left(-\dfrac{l{+}k}{\rho\left(\frac{1}{\mu}{-}k\right)}\right)\left(\dfrac{\frac{1}{\mu}{-}k}{l{+}\frac{1}{\mu}}\right)^{i{+}1{-}l}.
\end{align}

2) If $l>i+1-\tfrac{1}{\mu}$, we have $(i+1-l)x > \tfrac{x}{\mu}-\tfrac{1}{\rho}$ when $x<\tfrac{1}{\rho\left(\tfrac{1}{\mu}{-}i{-}1{+}l\right)}$ and $(i+1-l)x < \tfrac{x}{\mu}-\tfrac{1}{\rho}$ when $x>\tfrac{1}{\rho\left(\tfrac{1}{\mu}{-}i{-}1{+}l\right)}$. By combining \eqref{Appendix_eq1} and \eqref{Appendix_eq3}, $y_i^{(l)}$ can be written as
\begin{align}\label{Appendix_eq5}
y_i^{(l)}&=S\cdot \displaystyle\int_{\frac{\mu}{\rho}}^{\frac{1}{\rho(1/\mu{-}i{-}1{+}l)}}\displaystyle\int_0^{\frac{x}{\mu}{-}\frac{1}{\rho}}\sum_{k=0}^{i+1-l}\binom{i+1-l}{k}(-1)^k\exp\left(-lx-z\right)(z-kx)^{i-l}U(z-kx)dzdx\notag\\
&+S \cdot\displaystyle\int_{\frac{1}{\rho(1/\mu{-}i{-}1{+}l)}}^\infty\displaystyle\int_0^{(i{+}1{-}l)x}\sum_{k=0}^{i+1-l}\binom{i+1-l}{k}(-1)^k\exp\left(-lx-z\right)(z-kx)^{i-l}U(z-kx)dzdx\notag\\
&=\dfrac{(i{+}1)!}{(i{+}1{-}l)!(l{-}1)!} \displaystyle\sum_{k=0}^{i{+}1{-}l} \binom{i{+}1{-}l}{k}(-1)^k \left(  \dfrac{\exp\left({-}\frac{k{+}l}{\rho(\frac{1}{\mu}{-}k)}\right)}{k{+}l}-\displaystyle\sum_{s=0}^{i{-}l}\left( \dfrac{(i{+}1{-}l{-}k)^s}{(i{+}1)^{s{+}1}}\cdot\right.\right. \notag \\
& \,\,Q\left(1{+}s,\dfrac{i{+}1}{\rho(\frac{1}{\mu}{-}i{-}1{+}l)}\right){+}\exp\left({-}\frac{k{+}l}{\rho(\frac{1}{\mu}{-}k)}\right)\dfrac{(\frac{1}{\mu}{-}k)^s}{(\frac{1}{\mu}{+}l)^{s+1}}{\cdot}\left(\vphantom{Q\left(1{+}s,{-}\dfrac{1}{\rho(i{+}1{-}l{-}\frac{1}{\mu})}{\cdot}\dfrac{(i{+}1{-}l{-}k)(\frac{1}{\mu}{+}l)}{\frac{1}{\mu}{-}k} \right)}1{-}\right.\notag \\
&\left.\left.\left.Q\left(1{+}s,{-}\dfrac{1}{\rho(i{+}1{-}l{-}\frac{1}{\mu})}{\cdot}\dfrac{(i{+}1{-}l{-}k)(\frac{1}{\mu}{+}l)}{\frac{1}{\mu}{-}k} \right)\right)   \right) \vphantom{\dfrac{\exp\left({-}\frac{k{+}l}{\rho(\frac{1}{\mu}{-}k)}\right)}{k{+}l}} \right),
\end{align}
where $Q(s,t)=\frac{1}{(s-1)!}\int_t^\infty e^{-x}x^{s-1}dx$ is the regularized upper incomplete gamma function.
By combining \eqref{Appendix_eq4} and \eqref{Appendix_eq5}, we can obtain \eqref{y_i^l}.

With $\mu \geq 1$, for $l=1,..., i$, we have $l\leq i+1-\frac{1}{\mu}$. By substituting $\left\lceil1/\mu\right\rceil{-}1=0$ into \eqref{Appendix_eq4}, we have
\begin{equation}\label{Appendix_eq6}
y_i^{(l)}=\binom{i{+}1}{l}\exp\left(-\tfrac{l\mu}{\rho}\right)\cdot\left(\tfrac{1}{l\mu+1}\right)^{i{+}1-l}.
\end{equation}
For $l=i+1$, we have $l>i+1-\frac{1}{\mu}$. By substituting $i+1-l=0$ into \eqref{Appendix_eq5}, we have
\begin{equation}\label{Appendix_eq7}
y_i^{(i+1)}=\exp\left(-\tfrac{(i+1)\mu}{\rho}\right).
\end{equation}
\eqref{Mu>1} can be obtained by combining \eqref{Appendix_eq6} and \eqref{Appendix_eq7}.

\section{Derivation of \eqref{Corollary_smallthreshold}}\label{Proof_Corollary_smallthreshold}

1) For unordered SIC, by combining \eqref{ric} and \eqref{rij_sn}, we have
\begin{equation}\label{Corollary_smallthreshold_ri_unordered}
r_{n-1}^{NS\_l}|_{\mu=\tfrac{1}{n}}=\left(1+\tfrac{\exp\left(-\tfrac{1}{n\rho}\right)}{n\left(1+\tfrac{1}{n}\right)^{n-1}}\right)^n-1.
\end{equation}
For large $n$ and $\rho$, we have $\exp\left(-\frac{1}{n\rho}\right)\approx 1$ and $\left(1+\frac{1}{n}\right)^{n-1} \approx e$. \eqref{Corollary_smallthreshold_ri_unordered} can then be approximated to $r_{n-1}^{NS\_l}|_{\mu=\frac{1}{n}}\stackrel{\text{for large } n,\; \rho}{\approx}\left(1+\tfrac{e^{-1}}{n}\right)^n-1\stackrel{\text{for large } n}{\approx}e^{e^{-1}}-1$.

2) For ordered SIC, according to \eqref{rij_so}, we have
\begin{equation}\label{Corollary_smallthreshold_ordered 1}
r_{n-1}^{OS}=\tfrac{1}{n}\sum_{m=1}^{n}\Pi_{l=1}^m y_{n-1}^{(l)}.
\end{equation}
With $\mu=\frac{1}{n}$, $i+1-\tfrac{1}{\mu} \leq 0$ for $i=0,..., n-1$. Therefore, according to \eqref{y_i^l}, we have
\begin{align}\label{Corollary_smallthreshold_ordered_2}
y_{n-1}^{(l)}|_{\mu=\tfrac{1}{n}}&=\tfrac{n!}{(n{-}l)!(l{-}1)!} \displaystyle\sum_{k=0}^{n{-}l} \binom{n{-}l}{k}(-1)^k \left(  \tfrac{\exp\left({-}\tfrac{k{+}l}{\rho(n{-}k)}\right)}{k{+}l}-\displaystyle\sum_{s=0}^{n{-}l{-}1}\left( \tfrac{(n{-}l{-}k)^s}{n^{s{+}1}}Q\left(1{+}s,\tfrac{n}{\rho l}\right)\right.\right.  \notag\\
&{+}\left.\left.\exp\left({-}\tfrac{k{+}l}{\rho(n{-}k)}\right)\tfrac{(n{-}k)^s}{(n{+}l)^{s+1}}{\cdot}\left(1{-}Q\left(1{+}s,\tfrac{1}{\rho l}{\cdot}\tfrac{(n{-}l{-}k)(n{+}l)}{n{-}k} \right)\right)   \right) \vphantom{\tfrac{\exp\left({-}\frac{k{+}l}{\rho(n{-}k)}\right)}{k{+}l}} \right),
\end{align}
for $l=1,..., n$. Note that for large $\rho$, $\exp\left({-}\frac{k{+}l}{\rho(n{-}k)}\right) \approx1$, and the item
\begin{equation*}\label{Corollary_smallthreshold_ordered_3}
\displaystyle\sum_{s=0}^{n{-}l{-}1}\left( \tfrac{(n{-}l{-}k)^s}{n^{s{+}1}}Q\left(1{+}s,\tfrac{n}{\rho l}\right)+\exp\left({-}\tfrac{k{+}l}{\rho(n{-}k)}\right)\tfrac{(n{-}k)^s}{(n{+}l)^{s+1}}{\cdot}\left(1{-}Q\left(1{+}s,\tfrac{1}{\rho l}{\cdot}\tfrac{(n{-}l{-}k)(n{+}l)}{n{-}k} \right)\right)   \right)
\end{equation*}
can be approximated to $\sum_{s=0}^{n{-}l{-}1} \tfrac{(n{-}l{-}k)^s}{n^{s{+}1}}=\left(1-\left(\tfrac{n-l-k}{n}\right)^{n-l}\right)/(k+l)$ because $Q(a,b)$ approaches $1$ as $b$ approaches $0$. Therefore, $y_{n-1}^{(l)}|_{\mu=\frac{1}{n}}$ for $l=1,..., n$ can be approximated to
\begin{align}\label{Corollary_smallthreshold_ordered_4}
y_{n-1}^{(l)}|_{\mu=\tfrac{1}{n}}\stackrel{\text{for large } \rho}{\approx} \tfrac{n!}{(n{-}l)!(l{-}1)!} \displaystyle\sum_{k=0}^{n{-}l} \binom{n{-}l}{k} \tfrac{(-1)^k}{k{+}l}\left(\tfrac{n-l-k}{n}\right)^{n-l}= 1.
\end{align}
By combining \eqref{Corollary_smallthreshold_ordered 1} and \eqref{Corollary_smallthreshold_ordered_4}, we can obtain that $r_{n-1}^{OS}\Large|_{\mu=\frac{1}{n}}\approx 1$ for large $\rho$.


\section{Derivation of \eqref{Corollary_largethreshold_ri_unordered}} \label{Proof_Corollary_largethreshold_ri_unordered}
1) For ordered SIC, $r_i^{OS}$ can be written as
\begin{equation}\label{Corollary_largethreshold_ri_ordered_1}
r_i^{OS}=\tfrac{1}{i+1}\left(y_i^{(1)}+y_i^{(1)}y_i^{(2)}+y_i^{(1)}y_i^{(2)}y_i^{(3)}+\cdots+y_i^{(1)}\cdots y_i^{(i+1)}\right),
\end{equation}
according to \eqref{rij_so}. For $\mu>1$, we have $\tfrac{y_i^{(1)}}{i+1}=r_i^C$ by combining \eqref{Mu>1} and \eqref{ric}. Therefore, we have
\begin{equation}\label{Corollary_largethreshold_ri_ordered_2}
\lim_{\mu\rightarrow\infty}\tfrac{r_i^{OS}}{r_i^C}=\lim_{\mu\rightarrow\infty} 1+y_i^{(2)}+y_i^{(2)}y_i^{(3)}+\cdots+y_i^{(2)}\cdots y_i^{(i+1)}.
\end{equation}
As $\mu\rightarrow\infty$, it can be easily obtained from \eqref{Mu>1} that
$\lim_{\mu\rightarrow\infty} y_i^{(l)}=0$.
We can then obtain from \eqref{Corollary_largethreshold_ri_ordered_2} that $\lim_{\mu\rightarrow\infty}\tfrac{r_i^{OS}}{r_i^C}=1$.

2) For unordered SIC, we have
\begin{equation}\label{Proof_Corollary_captureratio_unordered_eq2}
\lim_{\mu\rightarrow\infty}\tfrac{r_i^{NS\_l}}{r_i^C}=\lim_{\mu\rightarrow\infty}\tfrac{(1+\mu r_i^C)^{i+1}-1}{(i+1)\mu r_i^C},
\end{equation}
according to \eqref{rij_sn}. Further note from \eqref{ric} that
\begin{equation}\label{Proof_Corollary_captureratio_unordered_eq1}
\lim_{\mu\rightarrow\infty} \mu r_i^C =\lim_{\mu\rightarrow\infty} \tfrac{\rho}{\exp\left(\tfrac{\mu}{\rho}\right)\left(1+\mu\right)^{i+1}\left(1+\mu+i\rho\right)}=0.
\end{equation}
We then have $\lim_{\mu\rightarrow\infty}\tfrac{r_i^{NS\_l}}{r_i^C}=1$ by combining \eqref{Proof_Corollary_captureratio_unordered_eq1} and \eqref{Proof_Corollary_captureratio_unordered_eq2}.

\section{Derivation of (\ref{maxThroughput})-(\ref{mu_0})} \label{Proof_maxThroughput}
According to \eqref{throughput_K0}, the first-order derivative of $\hat{\lambda}_{out}$ with respect to $q_0$ can be written as
\begin{equation}\label{first_order_throughput}
\tfrac{d \hat{\lambda}_{out}}{d q_0}=n\sum_{i=0}^{n-1}\binom{n-1}{i} r_i \left(1-q_0\right)^{n-2-i} q_0^i(1+i-nq_0).
\end{equation}
It can be easily obtained from \eqref{first_order_throughput} that
\begin{equation}\label{first_order_throughput_1}
\tfrac{d \hat{\lambda}_{out}}{d q_0}\mid_{q_0=0}=(n-1)r_0>0,\;\text{and} \;\; \tfrac{d \hat{\lambda}_{out}}{d q_0}\mid_{q_0=1}=nr_{n-1}-(n-1)r_{n-2}.
\end{equation}

\begin{figure*}[!tp]
\centering
\subfloat[]{
\label{firstorder_throughput_1}
\includegraphics[width=3.5in,height=2.4in]{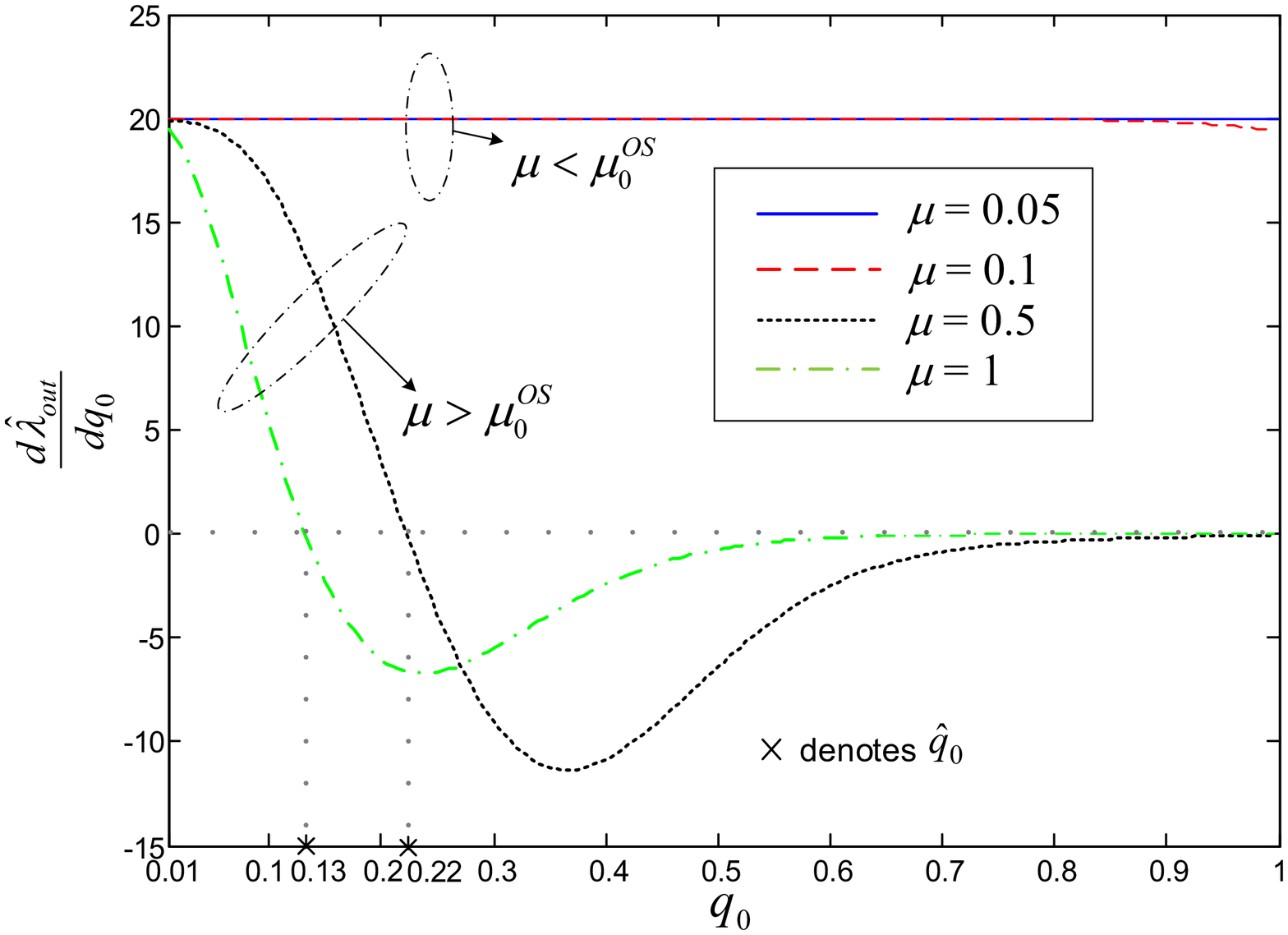}}
\hfil
\subfloat[]{
\label{firstorder_throughput_2}
\includegraphics[width=3.5in,height=2.4in]{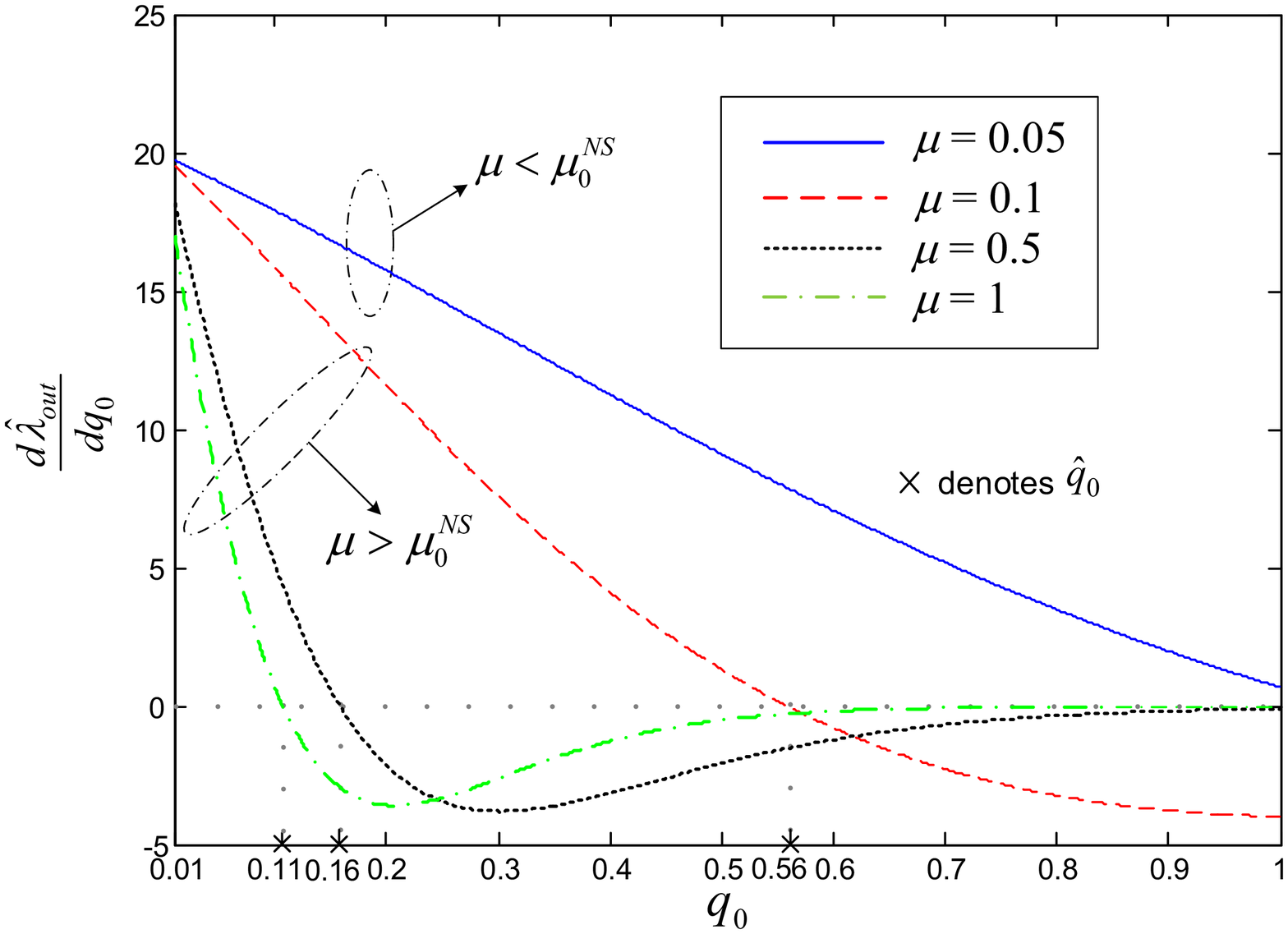}}
\caption{$\tfrac{d \hat{\lambda}_{out}}{d q_0}$ versus transmission probability $q_0$. $n=20$ and $\rho=20$dB. (a) Ordered SIC. (b) Unordered SIC.}
\label{firstorder_throughput}
\end{figure*}

Let $\mu_0$ denote the root of $nr_{n-1}-(n-1)r_{n-2}=0$. By substituting \eqref{rij_so} and \eqref{rij_sn} into \eqref{first_order_throughput}, we can obtain $\tfrac{d \hat{\lambda}_{out}}{d q_0}$ in the ordered SIC and unordered SIC cases, respectively, and Fig. \ref{firstorder_throughput} illustrates how $\frac{d \hat{\lambda}_{out}}{d q_0}$ varies with the transmission probability $q_0$. We can see from Fig. \ref{firstorder_throughput} that with $\mu<\mu_0$, $\tfrac{d \hat{\lambda}_{out}}{d q_0}>0$ for $q_0\in(0,1)$. Therefore, the maximum network throughput $\hat{\lambda}_{\max}$ is achieved at $q_0^*=1$, and we have $\hat{\lambda}_{\max}=n r_{n-1}$ according to \eqref{throughput_K0}. On the other hand, with $\mu \geq \mu_0$, we have $\frac{d \hat{\lambda}_{out}}{d q_0}\Large\mid_{q_0=1}\leq 0$. It can be clearly seen from Fig. \ref{firstorder_throughput} that $\tfrac{d \hat{\lambda}_{out}}{d q_0}> 0$ for $q_0\in(0,\hat{q}_0)$ and $\tfrac{d \hat{\lambda}_{out}}{d q_0}\leq 0$ for $q_0\in[\hat{q}_0,1]$, where $\hat{q}_0$ is the root of $\tfrac{d \hat{\lambda}_{out}}{d q_0}=0$. In this case, the maximum network throughput $\hat{\lambda}_{\max}$ is achieved at $q_0^*=\hat{q}_0$, and we have $\hat{\lambda}_{\max}=n\sum_{i=0}^{n-1}\binom{n-1}{i} r_i \left(1-\hat{q}_0\right)^{n-1-i} \hat{q}_0^{i+1}$ according to \eqref{throughput_K0}.

\begin{figure*}[!tp]
\centering
\subfloat[]{
\label{Mu_plot}
\includegraphics[width=3.5in,height=2.4in]{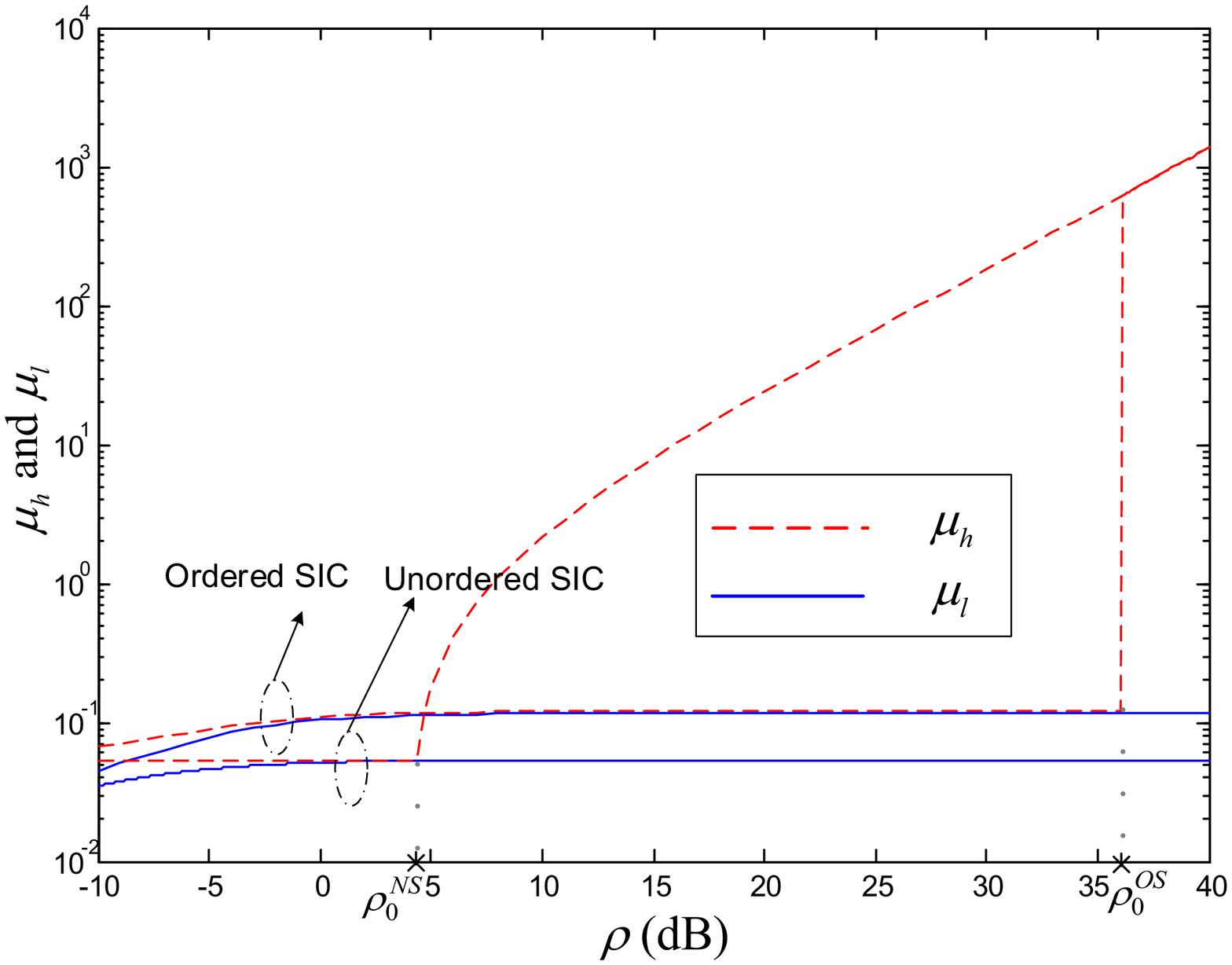}}
\hfil
\subfloat[]{
\label{f_plot}
\includegraphics[width=3.5in,height=2.4in]{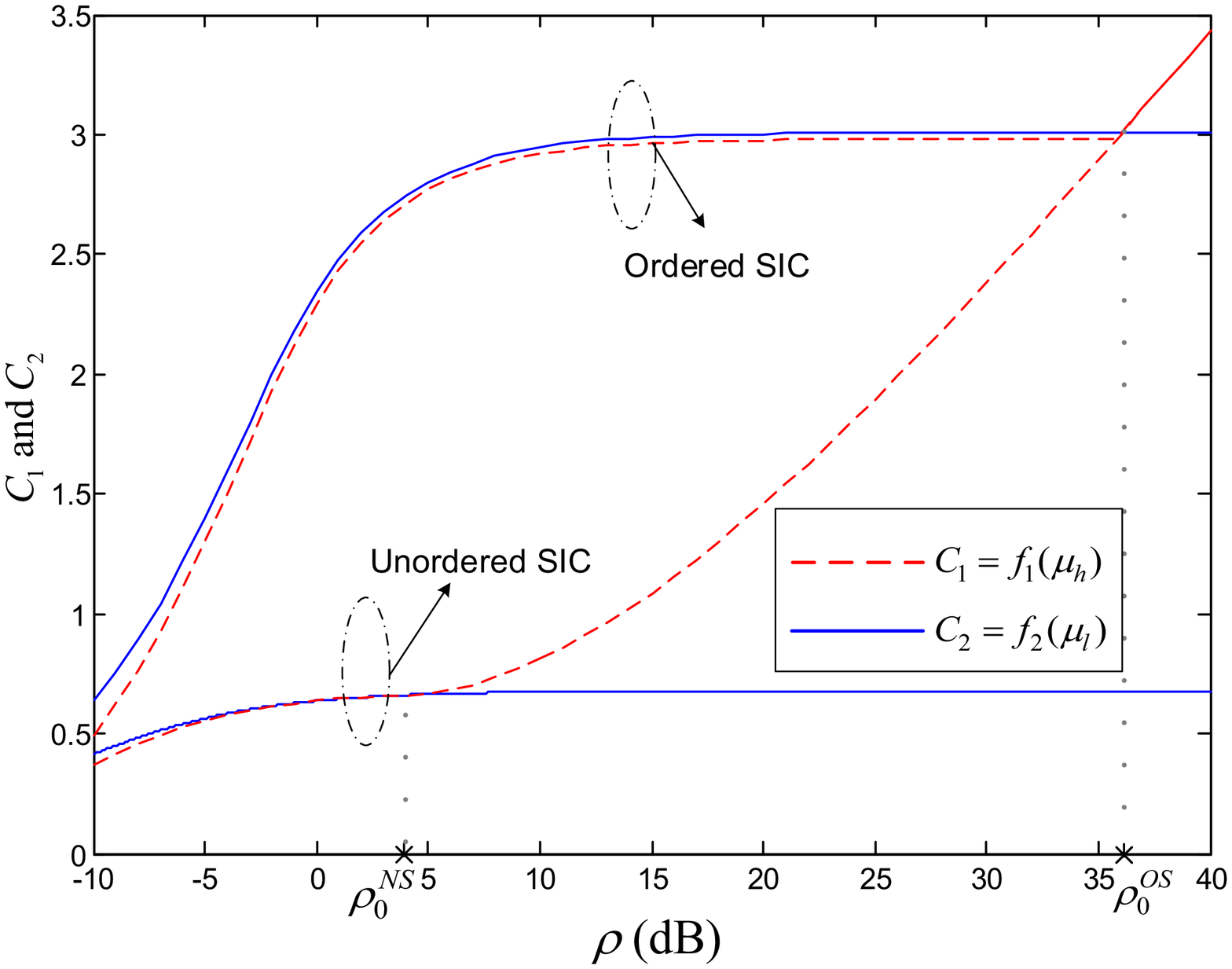}}
\caption{(a) $\mu_h$ and $\mu_l$ versus mean received SNR $\rho$. (b) $C_1$ and $C_2$ versus mean received SNR $\rho$. $n=20$.}
\label{plot}
\end{figure*}

\section{Derivation of (\ref{C_expression})-(\ref{optimal_threshold_expression})}\label{Derivation of C}

By substituting \eqref{rij_so} and \eqref{rij_sn} into \eqref{f_1_expre} and \eqref{f_2_expre}, respectively, we can numerically calculate $\mu_h$ and $\mu_l$ for both the ordered SIC and unordered SIC cases as shown in Fig. \ref{Mu_plot}, and further obtain $C_1=f_1(\mu_h)$ and $C_2=f_2(\mu_l)$ as shown in Fig. \ref{f_plot}. We can see from Fig. \ref{f_plot} that in both cases, $f_1(\mu_h)$ and $f_2(\mu_l)$ are monotonically increasing functions with respect to the mean received SNR $\rho$, and there exists a unique point $\rho_0$ such that when $\rho<\rho_0$, $f_1(\mu_h)<f_2(\mu_l)$ and $f_1(\mu_h)>f_2(\mu_l)$ when $\rho>\rho_0$, where $\rho_0$ is the root of $f_1(\mu_h)=f_2(\mu_l)$. We can then conclude that the maximum sum rate $C=C_1=f_1(\mu_h)$ if $\rho\geq\rho_0$, which is achieved when $\mu^*=\mu_h$. Otherwise, the maximum sum rate $C=C_2=f_2(\mu_l)$, which is achieved when $\mu^*=\mu_l$.

\section{Derivation of (\ref{maximum sum rate_collision})-(\ref{maximum throughput_collision})} \label{Aloha_collision-model}
Based on the collision model, a packet transmission is successful if and only if there are no concurrent transmissions and its received SNR is above the threshold $\mu$. The steady-state probability of successful transmission of HOL packets, $p$, can be then written as
\begin{equation}\label{collision-model_pA}
p=\text{Pr\{no concurrent packet transmissions\}}\cdot \text{Pr\{received SNR is above the threshold $\mu$\}}.
\end{equation}
According to \eqref{Probability-of-success_2}, the probability that there are no concurrent transmissions is given by
\begin{equation}\label{collision-model_pA1}
\text{Pr\{no concurrent packet transmissions\}}=\left(1-\tfrac{\pi_T}{p}\right)^{n-1}.
\end{equation}
Since the received SNR is exponentially distributed with mean $\rho$, the probability that the received SNR is above the threshold $\mu$ is given by
\begin{equation}\label{collision-model_pA2}
\text{Pr\{received SNR is above the threshold $\mu$\}}=\exp \left (-\tfrac{\mu}{\rho} \right ).
\end{equation}
By substituting \eqref{collision-model_pA1} and \eqref{collision-model_pA2} into \eqref{collision-model_pA}, we have
\begin{equation}\label{collision-model_pA3}
p=\left(1-\tfrac{\pi_T}{p}\right)^{n-1}\cdot \exp \left (-\tfrac{\mu}{\rho} \right )\stackrel{\text{for large}\; n}{\approx} \exp \left (-\tfrac{\mu}{\rho}-\tfrac{n\pi_T}{p} \right ),
\end{equation}
which can be further written as
\begin{equation}\label{collision-model_pA4}
p=\exp \left (-\tfrac{\mu}{\rho} -\tfrac{n}{\sum_{i=0}^{K-1}\tfrac{p\left(1-p\right)^i}{q_i}+\tfrac{\left(1-p\right)^K}{q_K}}\right ),
\end{equation}
according to \eqref{pi0}. With $q_i=q_0$, $i=0,...,K$, the fixed-point equation \eqref{collision-model_pA4} has a single non-zero root $p_A^c$, which is given by
$p_A^c=\exp \left(-\tfrac{\mu}{\rho}- n q_0 \right)$.
The network throughput in saturated conditions can be then obtained as $\hat{\lambda}_{out}^{collision}=n q_0 p_A^c=n q_0 \exp \left(-\tfrac{\mu}{\rho}- n q_0 \right)$, which is maximized at
\begin{equation}\label{collision-model_maxThroughput}
\hat{\lambda}_{\max}^{collision}=\exp\left(-1-\tfrac{\mu}{\rho}\right),
\end{equation}
when the transmission probability $q_0$ is set to be $q_{0}^{*,collision}=\tfrac{1}{n}$. By combining \eqref{maxRate_1} and \eqref{collision-model_maxThroughput}, the maximum sum rate can be written as
\begin{equation}\label{collision-model_maxRate}
C^{collision}=\max_{\mu>0}\exp\left (-1-\tfrac{\mu}{\rho}\right)\cdot \log_2(1+\mu).
\end{equation}
Let $f(\mu)$ denote the objective function of \eqref{collision-model_maxRate}. It can be easily shown that $f'(\mu)\geq 0$ for $\mu\in(0,\mu^{*,collision}]$ and $f'(\mu)< 0$ for $\mu\in(\mu^{*,collision},\infty)$, indicating that $f(\mu)$ has one global maximum at $\mu^{*,collision}$, where $\mu^{*,collision}=e^{{\mathbb W}_{0}(\rho)}-1$ is the root of $(\mu+1)^{(\mu+1)/{\rho}}=e$. 
\eqref{maximum sum rate_collision} and \eqref{maximum throughput_collision} can be then obtained by substituting $\mu^{*,collision}$ 
into \eqref{collision-model_maxRate} and \eqref{collision-model_maxThroughput}, respectively.

\end{document}